\pdfoutput=1
\documentclass[aps,prx,reprint,superscriptaddress,nofootinbib,floatfix]{revtex4-2}

\usepackage{amsmath,amssymb}
\usepackage{graphicx}
\usepackage{xcolor}
\usepackage{booktabs}
\usepackage[colorlinks=true,linkcolor=blue!50!black,citecolor=blue!50!black,urlcolor=blue!50!black]{hyperref}

\newif\ifLineNumbers
\LineNumbersfalse
\ifLineNumbers\usepackage{lineno}\AtBeginDocument{\linenumbers}\fi

\graphicspath{{figures/}}

\newif\ifDraftCallouts
\DraftCalloutsfalse
\newcounter{crem}
\definecolor{cremBG}{RGB}{253,246,231}
\definecolor{cremFR}{RGB}{176,122,48}

\definecolor{kfBlue}{RGB}{0,90,160}

\newcommand{\lna}{\ln\!A}
\newcommand{\mlna}{\langle\lna\rangle}
\newcommand{\gev}{\,\ensuremath{\mathrm{GeV}}}
\newcommand{\gv}{\,\ensuremath{\mathrm{GV}}}
\newcommand{\GSF}{GSF}

\newcommand{\gsfChiSq}{1261}

\newcommand{\gsfNdof}{972}

\newcommand{\gsfRedChiSq}{1.30}

\newcommand{\gsfRedChiSqShort}{1.3}

\newcommand{\gsfChiSqCorr}{805}

\newcommand{\gsfRedChiSqCorr}{0.83}

\newcommand{\gsfChiSqCorrEpos}{799}

\newcommand{\gsfNdofEpos}{968}

\newcommand{\gsfRedChiSqCorrEpos}{0.83}

\newcommand{\gsfNumEscales}{15}

\newcommand{\gsfEscaleGrapes}{0.89}

\newcommand{\gsfEscaleZGrapes}{-0.4}

\newcommand{\gsfEscaleIceCube}{1.04}

\newcommand{\gsfEscaleIceTopLowE}{1.04}

\newcommand{\gsfEscaleLhaaso}{1.00}

\newcommand{\gsfEscaleAuger}{0.87}

\newcommand{\gsfEscaleTA}{0.93}

\newcommand{\gsfEscaleTunka}{0.96}

\newcommand{\gsfEscaleTAtoAugerPct}{6}

\begin{document}

\title{Global Spline Fit: A unified data-driven view of the cosmic-ray
spectrum and mass composition from GeV to the highest energies}

\author{Anatoli Fedynitch}
\email[Corresponding author: ]{anatoli@sinica.edu.tw}
\affiliation{Institute of Physics, Academia Sinica, 11529 Taipei, Taiwan}
\author{Kozo Fujisue}
\affiliation{Institute of Physics, Academia Sinica, 11529 Taipei, Taiwan}
\author{Hans Dembinski}
\affiliation{Fakult\"at Physik, Technische Universit\"at Dortmund, 44221 Dortmund, Germany}
\author{Ralph Engel}
\affiliation{Karlsruhe Institute of Technology, D-76021 Karlsruhe, Germany}

\date{\today}

\begin{abstract}
The energy spectrum and mass composition of cosmic rays are measured by two
complementary classes of instruments: space- and balloon-borne detectors that
resolve individual elements up to sub-PeV energies, and ground-based air-shower
observatories that extend the reach to beyond $10^{11}$\gev{} but resolve only
broad mass groups. The Global Spline Fit (\GSF{}) is a data-driven
model that combines both into a single, self-consistent description of the
flux of all elements from hydrogen to nickel. The flux of four leading mass
groups is parametrized by cubic basis splines, imposing smoothness but no
astrophysical expectation on the spectral shape, while the energy-scale
offsets of the participating experiments are cross-calibrated within their
quoted systematic uncertainties as part of the fit. The model is defined at
the local interstellar spectrum, with the effect of solar modulation at the
lowest energies accounted for in the fit. The fit uses the most precise
recent data, among them the elemental spectra from AMS-02, CALET, and DAMPE,
the knee-region measurements of LHAASO, and the fluorescence-based
composition of the Pierre Auger Observatory, and describes about one thousand
data points with $\chi^2/\mathrm{ndf}\approx \gsfRedChiSqShort$, or $\gsfRedChiSqCorr$ after
de-weighting localized disagreements between data sets, demonstrating that
the global body of cosmic-ray data is consistent once energy scales are
aligned. The
model delivers the flux, the mass composition, and their full covariance,
along with a compact reduced representation for uncertainty propagation. We
further provide the cosmic-ray nucleon flux, which differs from
widely used parametrizations by 20--50\% over four decades in energy. This
exceeds the remaining model uncertainty several times over and has direct
consequences for atmospheric neutrino and muon flux predictions.
\end{abstract}

\maketitle

\section{Introduction}
\label{sec:intro}

The sources of the cosmic radiation, its mass composition, and its transport
through the Galaxy remain open questions~\cite{Gabici:2019jvz}, and no first-principles calculation
predicts the cosmic-ray flux at Earth with meaningful precision. Any accurate,
full-range description of the spectrum must therefore be constructed from
data. The measurements come in two distinct classes. Below a few hundred TeV,
cosmic rays are observed directly by magnetic spectrometers and calorimeters
flown on satellites and balloons, which identify individual elements, at
the lowest energies even isotopes, with percent-level precision. At higher
energies the steeply falling flux makes direct detection impractical, and
observations rely on extensive air showers, recorded through Cherenkov and
fluorescence light or through surface arrays that sample the electromagnetic
and muonic shower footprint. Air-shower observables constrain the primary
mass only to about one unit in the natural logarithm of the mass number,
$\lna$, so indirect experiments report the
fluxes of a few broad mass groups, or moments of the mass distribution such
as the mean logarithmic mass $\mlna$, rather than elemental spectra.

Any global description of the cosmic-ray flux must therefore bridge
elemental and group-level information in a consistent way. The Global Spline
Fit (\GSF{}) does this by decomposing the all-particle flux into four leading
mass groups (protons, helium, the oxygen group, and the iron group) that
span roughly equal intervals in $\lna$. In the energy range of
direct measurements the fit follows the individual elemental spectra and sums
them into the groups; at higher energies the same four group fluxes are
constrained by air-shower data. Both regimes are fit simultaneously with one
model.

The defining choice of the \GSF{} is its parametrization. Each leading flux
is a cubic basis spline in the logarithm of magnetic rigidity, which imposes
smoothness but no functional form motivated by acceleration or propagation
physics. This sets the \GSF{} apart from the model parametrizations in
common use; examples are the poly-gonato model~\cite{Hoerandel:2002yg}, the
H3a and H4a models of Gaisser~\cite{Gaisser:2011klf}, the models of
Gaisser, Stanev, and Tilav~\cite{Gaisser:2013bla}, of Gaisser and
Honda~\cite{Gaisser:2002jj}, of Zatsepin and
Sokolskaya~\cite{Zatsepin:2006ci}, and, more recently, the LHAASO-anchored
model of Ref.~\cite{Lv:2024wrs}. All of these describe the flux as a sum of
broken power laws with rigidity-dependent cutoffs. Over the past decade the
precision of direct measurements has revealed successive spectral breaks in
individual elemental fluxes that such rigid forms do not anticipate: a fixed
number of power-law segments under-represents the structure present in the
data and imprints the modeler's expectations on the result. A spline basis
with a full covariance, including the correlations between mass groups,
follows any feature the data require and represents the allowed range of
fluxes and compositions more faithfully. The goal of the \GSF{} is thus
deliberately descriptive: to represent the current observational knowledge of
the cosmic-ray flux, that is, the ensemble mean of the experiments and its
uncertainty, rather than to explain its origin.

Such a reference serves several goals. Atmospheric lepton fluxes at
high energies, the dominant background and calibration signal for neutrino
telescopes such as IceCube~\cite{IceCube:2016zyt},
KM3NeT~\cite{KM3Net:2016zxf}, Baikal-GVD~\cite{Baikal-GVD:2021zsq}, and the
planned P-ONE~\cite{P-ONE:2020ljt} and TRIDENT~\cite{TRIDENT:2022hql}
detectors, are computed from the
cosmic-ray nucleon flux, and their uncertainties inherit directly from it.
Air-shower experiments and studies of hadronic interactions, for instance of
the long-standing muon excess~\cite{Albrecht:2021cxw}, require a common
primary-flux reference with quantified errors.
Gamma-ray observatories treat cosmic rays as their dominant background.
Finally, because most experiments are not directly sensitive to the primary
mass, the \GSF{} composition can serve as an informative prior in analyses of
individual experiments, such as the LHAASO helium
spectrum~\cite{LHAASO:2025mlf}, the IceCube proton
fraction~\cite{IceCube:2024ouv}, or the IceCube measurement of the prompt
atmospheric neutrino flux~\cite{Abbasi:2025rmj}, much as global fits provide priors in
neutrino oscillation physics.

The first \GSF{} was shown in
2017~\cite{Dembinski:2017zsh}, with an intermediate update presented at
ICRC~2019~\cite{Schroder:2019agg} and more recent ones at subsequent
conferences~\cite{Fujisue:2025wnp,Dembinski:2025nmp}. The present analysis is
a significant rebuild on fully rewritten source code: it incorporates the data sets of the past decade, treats solar modulation, published covariances, and isotopic
sub-species explicitly, and ships as an open-source package with a reduced
representation for downstream error propagation. Section~\ref{sec:method}
describes the model and the fit, Sec.~\ref{sec:data} the data selection,
Sec.~\ref{sec:results} the results, Sec.~\ref{sec:discussion} the remaining
tensions and their interpretation, and Sec.~\ref{sec:nucleon} the derived
nucleon flux. Technical material (units and kinematics, the spline basis and
knot placement, the likelihood definitions and the detailed data treatment,
the model reduction, and fit diagnostics) is collected in the Supplemental
Material (SM).

\section{The global spline fit}
\label{sec:method}

\subsection{Flux model}
\label{sec:model}

We write the differential flux of a cosmic-ray species with charge $Z$ and
mass number $A$ as $J(E)=\mathrm{d}N/(\mathrm{d}E\,\mathrm{d}A\,\mathrm{d}t\,
\mathrm{d}\Omega)$. The natural variable of the fit is the magnetic rigidity
$R=pc/(Ze)$, in which acceleration and propagation in magnetic fields are
universal and in which solar modulation acts; conversions between $J(R)$ and
$J(E)$ are given in SM~S1.

\begin{figure}
\includegraphics[width=\columnwidth]{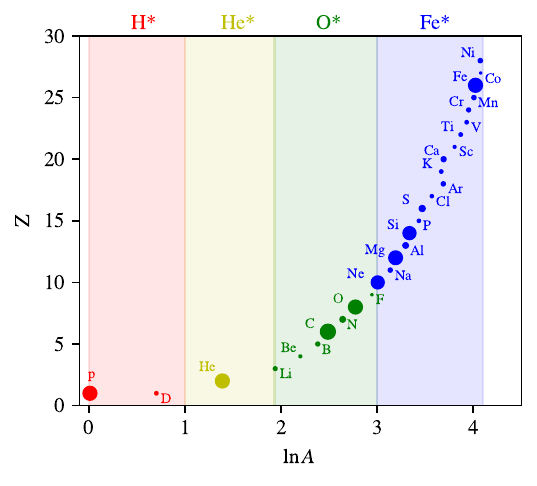}
\caption{Elements and mass groups of the \GSF{}. A group is named after its
leading element, marked by an asterisk; groups divide the $\lna$ axis into four
roughly equal parts. The area of each circle is proportional to the flux
ratio of the element to its group leader (values tabulated in the SM).}
\label{fig:massgroups}
\end{figure}

The elements from hydrogen to nickel are divided into four mass groups
covering roughly equal intervals in $\lna$, corresponding to the mass resolution of
air-shower experiments. Each group is named after its leading element, marked
with an asterisk: H$^{*}$, He$^{*}$, O$^{*}$, and Fe$^{*}$
(Fig.~\ref{fig:massgroups}). Four groups match the resolution of
high-energy experiments; a fifth would exceed what most experiments can
distinguish. The flux of each leading element $L$ is a clamped cubic basis
spline~\cite{DeBoor_splines} in $x=\ln(R/\mathrm{GV})$,
\begin{equation}
J_L(R) = \left(\frac{R}{\gv}\right)^{-3}\,\sum_k a_{Lk}\,b_k(x),
\label{eq:spline}
\end{equation}
where the $b_k$ are the cubic B-spline basis functions, the amplitudes
$a_{Lk}$ are free parameters, and the factor $R^{-3}$
removes the leading trend of the spectrum for numerical stability. The knots
are approximately uniform in $x$, with a higher density around prominent
spectral features such as the knee and the ankle. The choice of the knot
locations is a remaining degree of freedom of the \GSF{} parametrization;
the knot table, the placement rationale, and robustness tests are given in
SM~S1.
In the configuration used here, the leading and sub-leading splines
together comprise 321 amplitudes, of which 250 are free in the fit; the
remainder are held at zero where the non-negativity constraint of the
pre-fit is active. Together with the 15 energy-scale offsets of
Sec.~\ref{sec:escale} and the 9 solar-modulation shifts of
Sec.~\ref{sec:likelihood}, the fit has 274 free parameters in total.

Up to the highest energy of direct measurements, every sub-leading element is described
by its own short spline, so that the fit effectively interpolates the direct
elemental data. Above that energy, the flux of each sub-leading element follows the shape of its group
leader scaled by their measured abundance ratio. In previous \GSF{}
releases this ratio was held constant; here both its normalization and its
index are anchored to an error-weighted fit of the measured ratio over the
last decade of the element's data, and the ratio is extrapolated as a power
law in rigidity that saturates to a constant at $R = 5$\,PV (SM~S1),
capturing the
measured energy dependence of the sub-leading--to--leading ratios, most
notably the decline of the secondary
species~\cite{CALET:2025dgy,AMS:2018tbl}. The change has a percent-level effect
on the fit itself, entering only through the masses of the sub-leading
elements in the group sums, to which the air-shower observables are not
sensitive, but it matters for the extrapolated elemental fluxes.

\begin{figure}
\includegraphics[width=0.9\columnwidth]{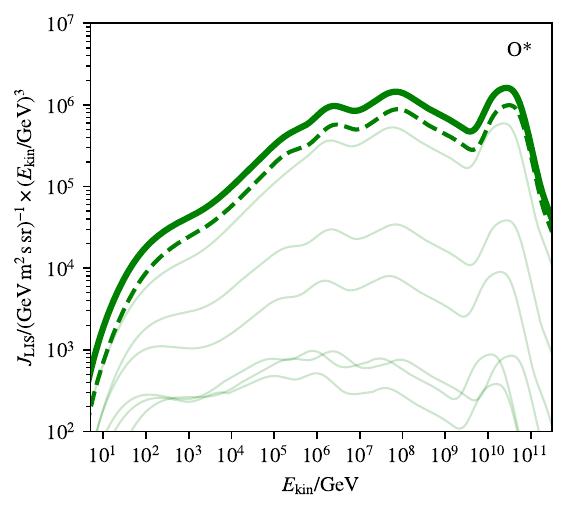}\\[2pt]
\includegraphics[width=0.9\columnwidth]{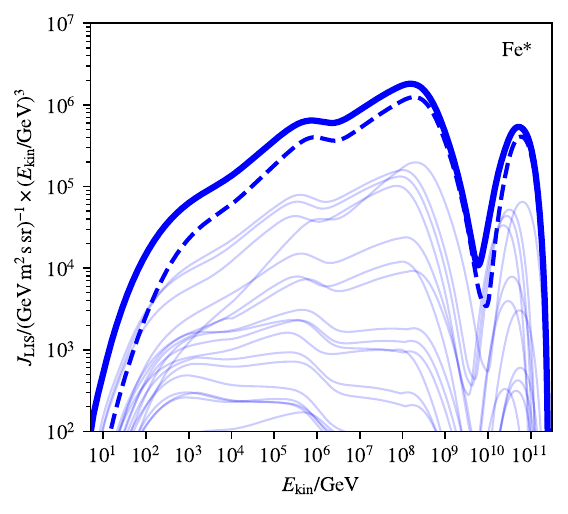}
\caption{Flux of the oxygen (top) and iron (bottom) groups (thick lines)
compared to the flux of the leading elements alone (dashed) and of the
sub-leading elements (thin lines). The offset between group and leading
element underlines the role of the sub-leading abundances.}
\label{fig:groups}
\end{figure}

The resulting group fluxes are substantially
larger than the fluxes of their leading elements alone
(Fig.~\ref{fig:groups}), which is why the distinction between elements and
groups is essential when direct and air-shower data are combined. As a new
element with respect to Ref.~\cite{Dembinski:2017zsh}, the AMS-02 deuteron
flux~\cite{AMS:2024idr} is carried as an explicit sub-leading $Z=1$ species
below the proton leader: D is the one case in which an isotope is both
measured separately and excluded from its element's flux (the AMS-02 proton
spectrum is deuteron-subtracted). The fitted ratio D/p rises from 0.025 to
0.026 across the AMS rigidity range and contributes $+0.018$ to the
$\mlna$ of the proton group.
No analogous treatment is applied to $^3$He: the AMS-02 helium flux
used in the fit contains both isotopes, and the measured $^3$He/$^4$He
ratio \emph{decreases} with rigidity~\cite{AMS:2019nij}, so that, unlike the
secondary-to-primary ratios discussed above, no growing $^3$He contribution
needs to be extrapolated. Treating the helium group as pure $^4$He
overestimates its $\mlna$ by at most a few per cent.

The all-particle flux is the sum over the groups and their members,
\begin{equation}
J(E) = \sum_L \sum_{j\in L} w_{Lj}\, J_L\bigl(R_j(E)\bigr)\,
\Bigl(\frac{\mathrm{d}R}{\mathrm{d}E}\Bigr)_{\!j},
\label{eq:total}
\end{equation}
where $w_{Lj}$ are the abundance weights and $R_j(E)$ the per-species
kinematic conversion from energy to rigidity (SM~S1); we write $J_{Lj}(E)$
for the individual terms of the sum, the flux of species $j$ per energy
interval. The same splines predict every composition
observable used in the fit: group fluxes, group fractions of the total, and
the mean logarithmic mass
\begin{equation}
\mlna(E) \;=\; \frac{\sum_L \sum_{j\in L} \ln\!A_j \, J_{Lj}(E)}
                    {\sum_L \sum_{j\in L} J_{Lj}(E)},
\label{eq:lna}
\end{equation}
with the sums running over all elements of all groups.

\subsection{Energy-scale cross-calibration}
\label{sec:escale}

No standard candle exists for air showers, and the absolute energy scale of
each air-shower experiment carries a correlated systematic uncertainty of
typically 10--20\%, which is usually the dominant uncertainty of flux data. For a steeply
falling spectrum even a modest scale shift produces a large apparent flux
change: if an experiment's energy scale is offset by a factor
$f_e=\tilde E/E$, the flux it reports is distorted as
\begin{equation}
\tilde J(\tilde E) \;=\; \frac{1}{f_e}\, J\!\left(\tilde E/f_e\right),
\label{eq:rescale}
\end{equation}
which for a power law $J\propto E^{-2.7}$ turns a $+10\%$ scale offset into
an 18\% flux excess.

The \GSF{} treats these offsets as nuisance parameters. Each experiment $e$
with a quoted energy-scale uncertainty $\sigma_{E,e}$ receives one parameter
$z_e$, entering the comparison of model and data through
$f_e = 1 + \sigma_{E,e}\, z_e$, and a Gaussian penalty $z_e^2$ is added to
the fit objective. The official energy-scale uncertainty of each
experiment is used where one is published; where none is available, a
representative value informed by the detection technique and by
cross-checks against neighboring experiments is adopted (marked by a
dagger in Table~\ref{tab:experiments}). The ensemble of experiments is thereby cross-calibrated
algorithmically, within the quoted systematics, with the percent-level
energy scales of the magnetic spectrometers acting as the absolute anchor.
The fitted offsets are themselves a result of the analysis and are
presented in Sec.~\ref{sec:results:scales}.

\subsection{Likelihood and data treatment}
\label{sec:likelihood}

Each data set enters a generalized least-squares objective with a covariance
built from the published statistical and systematic uncertainties, with a
default correlation of $\rho=0.5$ among the systematic uncertainties of the
data points within a set. Where
experiments publish covariance or correlation information we use it directly:
the IceCube mass-fraction correlations, the posterior correlations of the
Auger fluorescence-detector composition fractions, and the structural
anti-correlation between the LHAASO proton and helium spectra
(Sec.~\ref{sec:data}). Four classes of observables enter: elemental fluxes in
rigidity from direct experiments, group fluxes in energy and group fractions
from air-shower arrays, and $\mlna$. Their residual definitions
follow standard practice and are collected in SM~S2.

Direct measurements at low rigidity are modulated by the heliospheric
magnetic field. Rather than demodulating the data, the fit forward-models
solar modulation: each affected data set is compared to the local
interstellar spectrum (LIS) spline transformed by the force-field
approximation and averaged over that experiment's observation window,
\begin{equation}
\langle J\rangle(R) = \frac{1}{N}\sum_{t\,\in\,\mathrm{window}}
J_\mathrm{FF}\bigl(R;\phi_t\bigr),
\label{eq:solarmod}
\end{equation}
where $N$ is the number of monthly bins $t$ in the window,
using the monthly modulation potentials $\phi_t$ of the
Ghelfi--Maurin--Derome neutron-monitor
reconstruction~\cite{Ghelfi:2016pcv} over each experiment's observation
window (for long windows a dozen flux-weighted $\phi$ bins reproduce the
full monthly average to better than $0.1\%$); the average is linear in the
spline amplitudes and adds no cost to the fit. An alternative parameter set
fitted with the reconstruction of Usoskin
et~al.~\cite{Usoskin:2017cli} is provided alongside the default; the two
differ only in the interstellar spectrum below $\sim 10$\gv{} (SM~S2). This treatment matters at the percent level for
instruments with long or misaligned observation windows (the AMS-02 data
releases combine different windows per element) and replaces the single
per-experiment data demodulation used in Ref.~\cite{Dembinski:2017zsh}. The
uncertainty of the reconstructed potentials is propagated by fitted
nuisance shifts of $\phi(t)$: a common mode shared by all modulated data
sets, and a coherent per-experiment shift of each observation window. Both
are constrained by Gaussian priors and fitted alongside the energy-scale
offsets (SM~S2). Direct
data enter the fit down to $R=0.5$\gv{}; the lowest points are the
ACE-CRIS solar-minimum spectra, whose force-field mapping places them at
interstellar rigidities above $1.3$\gv{}, safely above the first spline
knots. Points below $0.5$\gv{} are dropped, where the force-field
approximation becomes unreliable. The published fluxes are
corrected for the geomagnetic cutoff by the experiments themselves, so no
cutoff treatment is required on the model side.

\subsection{Fitting procedure}
\label{sec:fitting}

The fit minimizes a generalized least-squares objective over the spline
amplitudes $\{a_{Lk}\}$ and the energy-scale parameters $\{z_e\}$,
\begin{equation}
\Phi = \sum_{d}\, \mathbf{r}_d^{\,T}\, V_d^{-1}\,
\mathbf{r}_d + \sum_{e} z_e^{2},
\label{eq:objective}
\end{equation}
where $\mathbf{r}_d$ is the residual vector of data set $d$ evaluated at the
shifted energy scale of its experiment, $V_d$ its covariance, and the second
term the Gaussian penalty that keeps each offset within its quoted
systematic ($z_e$ is the offset in units of $\sigma_{E,e}$,
Sec.~\ref{sec:escale}); the solar-modulation nuisances enter with analogous
unit-Gaussian penalties (residual definitions in SM~S2).

Equation~(\ref{eq:objective}) is not suited to a single-stage
gradient-descent minimization: at fixed
energy-scale offsets the model is linear in the spline amplitudes with a
positivity constraint, whereas the offsets enter nonlinearly; moreover,
where only all-particle data constrain the sum of the groups, amplitudes of
different mass groups can be traded against one another, and amplitudes
outside the rigidity range covered by data must be pinned at zero. The fit
therefore proceeds in two nested stages. First, a gradient-free search
(BOBYQA~\cite{bobyqa,nlopt}) explores the offsets; at every trial point the
several hundred spline amplitudes are solved exactly by non-negative least squares
(NNLS~\cite{nnls}), which enforces $J_L\ge 0$ and sets amplitudes outside
the range covered by data exactly to zero. Second, starting from this solution, all
remaining free parameters (the spline amplitudes not fixed at zero, the
energy-scale offsets, and the solar-modulation shifts) are minimized
jointly by a bounded Gauss--Newton iteration acting on the
covariance-whitened residuals (the residual vectors multiplied by a matrix
square root of the inverse covariance, so that the objective becomes a
plain sum of squares), which takes linearized bounded-variable
least-squares steps with a backtracking line search and converges within a
few tens of Jacobian evaluations; it replaces the MINUIT-based minimization
of previous releases. After the first convergence, the data are de-weighted
where the experiment ensemble disagrees beyond its stated errors, following
the scale-factor prescription of the Particle Data
Group~\cite{ParticleDataGroup:2026aaa}: in every bin whose reduced $\chi^2$
exceeds unity, all points in that bin are de-weighted by
$1/\sqrt{\chi^2_\mathrm{red}}$. The correction runs in two passes. The
first, as in previous \GSF{} releases, bins the directly measured
group-flux data per mass group in rigidity. The second pass, new in this
release, covers the observables the first cannot see: all-particle spectra,
mass fractions (including their published cross-fraction covariances), and
$\mlna$, binned together in total energy. A second minimization
with the corrected weights defines the final parameters. The parameter
covariance is obtained from derivatives of the analytic gradient at the
minimum and globally scaled by $\max(1,\chi^2_\mathrm{corr}/\mathrm{ndf})$,
so that post-correction over-consistency never shrinks the band. Quantitative
diagnostics of the correction, including the bins where it is applied, are
given in SM~S5.

\subsection{What the uncertainty band means}
\label{sec:covariance}

The \GSF{} uncertainty is constructed to \emph{cover} the spread among
experiments, including their energy-scale freedom. An estimator that shrinks
to the precision of the single most precise instrument would misstate what
the ensemble knows. The two-pass de-weighting and the one-sided global scaling
implement this for every fitted observable, whether flux, group fractions, or
$\mlna$: where experiments disagree beyond their stated errors, the
band widens accordingly (at the knee, the second knee, and in the
Auger--Telescope-Array energy range), and because the composition
observables participate in the correction they also gain weight in the
determination of the energy-scale offsets. Because statistical
and systematic uncertainties are combined, the resulting intervals do not
have strict frequentist coverage. They do, however, define a full parameter
covariance that propagates to any derived quantity through its Jacobian, so
that the tension of an external measurement or model prediction with the
\GSF{} can be quantified as a pull, the signed deviation from the band
divided by the combined uncertainty of the two. Amplitudes and
offsets form one joint covariance, so users of the published model can
propagate the energy-scale freedom together with the flux uncertainty; the
role of the global-scale mode is discussed in
Sec.~\ref{sec:discussion:scales}. Propagation recipes for derived quantities are given
in SM~S2, and a compact reduced representation of the covariance is
described in Sec.~\ref{sec:nucleon} and SM~S4.

\section{Data selection}
\label{sec:data}
\begin{figure}[!t]
\includegraphics[width=\columnwidth]{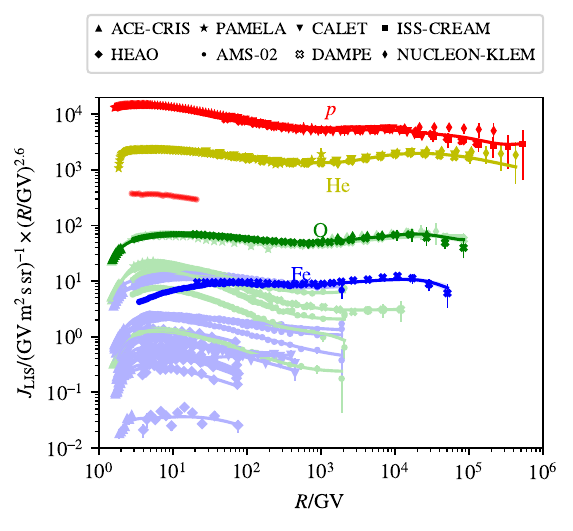}
\caption{Direct measurements of the four leading elements versus rigidity,
compared to the \GSF{} ($\pm1\sigma$ bands). Curves and data are both local
interstellar spectra: each experiment's points are demodulated with the
window-averaged modulation potential of its own observation period (SM~S2).
Unlabeled points show
sub-leading elements of the O$^{*}$ and Fe$^{*}$ groups; the deuteron
(dimmed) is carried as a sub-leading species of the proton group. Error bars
combine statistical and systematic uncertainties.}
\label{fig:direct}
\end{figure}

\begin{table*}
\caption{Data sets in the \GSF{}. For each experiment we list the detection
technique, the observables used, and the quoted energy-scale systematic
$\sigma_E/E$ that defines the offset parameter of
Eq.~(\ref{eq:rescale}). Spectrometer scales (---) are treated as exact.
Daggers ($\dagger$) mark $\sigma_E/E$ values that are not published by the
experiment but adopted here as representative conventions; widening these
from 10\% to 15\% has a negligible impact on the fit (all offsets move by
less than 0.015 in $f_e$ and the total $\chi^2$ is unchanged).
Detailed treatment (observation windows, hadronic models, covariance
structure) is tabulated in the SM.}
\label{tab:experiments}
\begin{ruledtabular}
\begin{tabular}{llllc}
Experiment & Technique & Observables in fit & $\sigma_E/E$ & Ref. \\
\colrule
\multicolumn{5}{l}{\emph{Direct (elemental fluxes vs.\ rigidity; solar modulation per observation window)}}\\
ACE-CRIS & satellite spectrometer & sub-leading elements; O & --- & \cite{vonRosenvinge:2013zqn,deNolfo:2006qj} \\
HEAO-3 & satellite & heavy elements P--Ni (w/o S, Fe) & 2\% & \cite{HEAO_data} \\
PAMELA & satellite spectrometer & p, He & --- & \cite{PAMELA:2011mvy} \\
AMS-02 & station spectrometer & p--Fe, D & --- & \cite{AMS:2021nhj,AMS:2024idr,AMS:2021lxc,AMS:2021brg,AMS:2023anq} \\
CALET & station calorimeter & p, He, Ni, Ti, Cr & 2\% & \cite{CALET:2022vro,CALET:2023nif,CALET:2022ajb,CALET:2025dgy} \\
DAMPE & satellite calorimeter & p, He, C, O, Fe, B & 2\% & \cite{DAMPE:2025opn,DAMPE:2024qwc} \\
ISS-CREAM & station calorimeter & p & 2\% & \cite{Choi:2022aht} \\
NUCLEON-KLEM & satellite & p, He, C, O; all-particle & 5\% & \cite{Gorbunov:2018stf} \\
\colrule
\multicolumn{5}{l}{\emph{Indirect (air showers: group fluxes, fractions, $\mlna$)}}\\
HAWC & surface array & all-particle; p+He & 16\% & \cite{HAWC:2021ubt,HAWC:2022zma} \\
GRAPES-3 & surface array & p & 25\%$^\dagger$ & \cite{GRAPES-3:2024mhy} \\
LHAASO & surface + imaging Cherenkov & all-particle; $\mlna$; p; He & 8\% & \cite{LHAASO:2024knt,LHAASO:2025mlf,Cao:2025nfo} \\
IceCube/IceTop & surface + in-ice array & all-particle; 4 fractions & 10\%$^\dagger$ & \cite{IceCube:2019hmk} \\
IceTop (low energy) & surface array & all-particle & 10\%$^\dagger$ & \cite{IceCube:2020yct} \\
Tunka-133 & non-imaging Cherenkov & all-particle; 4 fractions & 10\%$^\dagger$ & \cite{Budnev:2020oad,Prosin:2014dxa} \\
TALE (hybrid) & hybrid & $\mlna$ & 10\%$^\dagger$ & \cite{TelescopeArray:2026rdu} \\
Telescope Array & surface array & all-particle ($E\ge 10^{18.25}$\,eV) & 21\% & \cite{Ivanov:2020rqn} \\
Pierre Auger & hybrid & all-particle; 4 FD fractions & 14\% & \cite{PierreAuger:2021hun,PierreAuger:2026qbt} \\
\colrule
\multicolumn{5}{l}{\emph{Not considered: KASCADE-Grande~\cite{Kang:2023lre};
CALET B~\cite{CALET:2022dta}, C, O, Fe~\cite{CALET:2025dgy} (SM~S3);
H.E.S.S./VERITAS iron~\cite{HESS:2007rdt,VERITAS:2018gjd};}}\\
\multicolumn{5}{l}{\emph{\phantom{Not considered: }NUCLEON-KLEM iron; ACE-CRIS iron;
TALE monocular~\cite{TelescopeArray:2018bya}}}\\
\end{tabular}
\end{ruledtabular}
\end{table*}

Table~\ref{tab:experiments} lists the data entering the fit; several data
sets are accessed through the Cosmic-Ray Database~\cite{Maurin:2023alp}. The selection
follows the principles of Ref.~\cite{Dembinski:2017zsh}: we use the most
recent and most detailed release of each measurement, prefer results
interpreted with post-LHC hadronic interaction models wherever they exist,
favor techniques with smaller model dependence (optical and hybrid air-shower
observation, multi-component detection), and require composition sensitivity
at high energy. Exceptions are made where a gap in coverage would otherwise
open: the IceCube composition analysis is retained although its only
published interpretation is based on the pre-LHC model SIBYLL~2.1, and the
Telescope Array surface-detector spectrum is retained although it carries
no composition information, since it is the only measurement of the flux
at the highest energies from the northern hemisphere and complements
Auger. The Telescope Array fluorescence composition, by contrast, is not
included: its published interpretations rely on pre-LHC hadronic models,
and its statistical power is small compared to the Auger fluorescence
measurements.

Since 2017, when the first \GSF{} was presented, the observational landscape
has changed most on the direct side. AMS-02~\cite{AMS:2021nhj} provides percent-level elemental
spectra from hydrogen to iron; CALET~\cite{CALET:2022vro} and
DAMPE~\cite{DAMPE:2025opn} extend calorimetric elemental
measurements to $\sim$100~TV, with DAMPE now covering p, He, C, O, and Fe
and observing a common softening near 15~TV, so that direct, charge-resolved
data reach almost to the PeV scale (Fig.~\ref{fig:direct}). In the knee region, LHAASO contributes the
all-particle flux and $\mlna$ with a composition-independent energy
scale~\cite{LHAASO:2024knt} and the proton and helium spectra through the
knee~\cite{LHAASO:2025mlf}, which extend the earlier stand-alone
measurement of the proton knee~\cite{LHAASO:2025byy}; its absolute energy
scale is calibrated to 8\%
with the cosmic-ray Moon shadow~\cite{Cao:2025nfo}. The TALE hybrid analysis constrains $\mlna$
across the second-knee region~\cite{TelescopeArray:2026rdu}, and at the
highest energies the Auger composition is taken from the 17-year
fluorescence-detector analysis~\cite{PierreAuger:2026qbt}, which supersedes
the earlier neural-network surface-detector results that were calibrated to
it. IceCube spectra and mass fractions are used in their published form with
the full inter-group correlation matrix~\cite{IceCube:2019hmk}. Several of
these inputs carry explicit covariance information into the fit; for all
others the stated statistical and systematic uncertainties are used as
described in Sec.~\ref{sec:likelihood}.

Hadronic interaction models deserve a specific policy, since too few
air-shower data sets are interpreted with the \emph{same} model to allow
separate global fits per model. Where a collaboration publishes every
interpretation per energy bin and quotes the model term separately, as LHAASO
does, we adopt its own combination: the unweighted mean of the three
interpretations, with the published per-model systematic and half the model
spread added in quadrature. For the Auger composition,
where interpretations under SIBYLL-2.3e and EPOS-LHC-R differ strongly, the
delivered fit covers both: we fit separately under each interpretation and
combine the two at the covariance level (the mixture of
Sec.~\ref{sec:results:mixture}), so that the published uncertainty spans the
two hadronic interpretations rather than committing to one. SIBYLL-2.3e is the
baseline within the mixture: no other data set in the fit is interpreted with
EPOS-LHC-R, and that interpretation alone would leave a discontinuity against
the lower-energy composition measurements
(Sec.~\ref{sec:discussion:slna}).

A small number of data sets is deliberately not considered, for the reasons
stated here. The CALET boron~\cite{CALET:2022dta}, carbon,
oxygen, and iron
spectra~\cite{CALET:2025dgy} lie 15--24\% below the consistent
AMS-02/DAMPE/NUCLEON-KLEM ensemble of the same elements, with mean pulls of
$-2.8\sigma$ to $-4.5\sigma$ against the fitted fluxes. This normalization
discrepancy is far beyond the 2\% CALET energy-scale freedom; it is
documented by the collaboration for iron and is under investigation. The comparison is shown in SM~S3. The CALET titanium and chromium spectra from
the same release show no such offset, with mean pulls of $-0.2\sigma$ and
$-0.7\sigma$ against the fitted fluxes, and are used. The NUCLEON-KLEM iron spectrum is not considered for
the same reason: its normalization lies about 40\% below the DAMPE iron
measurement across its range, while the NUCLEON-KLEM spectra of the
lighter elements agree with the ensemble. At the lowest energies, the ACE-CRIS iron spectrum is in
tension with the AMS-02 iron flux where their rigidity coverages meet after
the solar-modulation mapping, and is not considered; we prefer to leave the
sub-GV iron range unconstrained rather than anchor the spline to a
spectrum that disagrees with the higher-rigidity data. Conversely, the
heavy elements between phosphorus and nickel are taken from HEAO-3,
which remains the only source for them until the corresponding AMS-02
spectra are published. The KASCADE-Grande composition~\cite{Kang:2023lre}
cannot be reconciled with the neighboring experiments: accommodating it
would require an energy-scale shift far beyond its uncertainty interval
($f_e\approx 0.74$), and even with that shift it remains incompatible with
them. The tension is local to this data set and, given the large
uncertainties, barely pulls the result; it is removed for the consistency of
the ensemble. The
iron spectra of the imaging air-Cherenkov telescopes
H.E.S.S.~\cite{HESS:2007rdt} and VERITAS~\cite{VERITAS:2018gjd} are
superseded by the direct DAMPE iron measurement, which now reaches
$\sim$1.5~PeV. The low-energy IceTop knee spectrum~\cite{IceCube:2020yct}
enters with its own energy-scale offset, independent of the IceCube
analysis of coincident surface and in-ice
events~\cite{IceCube:2019hmk} (different trigger, reconstruction, and years); the
resulting offset and the normalization tension it reveals in the transition
region are discussed in Sec.~\ref{sec:results:scales}. The TALE
monocular fluorescence spectrum~\cite{TelescopeArray:2018bya} is not considered because its energy
estimator is composition-dependent and tied to the early TALE $\mlna$
result, which is in tension with LHAASO.
The Telescope Array spectrum enters through the combined
TALE+surface-detector release~\cite{Ivanov:2020rqn}, of which only the
surface-detector part above $10^{18.25}$\,eV is used: below that energy the
combined spectrum is carried by the TALE fluorescence data not considered
above (the stitch is visible in the data themselves as a jump of the relative
statistical uncertainty from $\approx 8.5\%$ to $\approx
1.8\%$).

\section{Results}
\label{sec:results}

\subsection{Fit quality}
\label{sec:results:quality}

The fit describes the complete data set of Table~\ref{tab:experiments} with
$\chi^2=\gsfChiSq$ for \gsfNdof{} degrees of freedom before the weight correction
($\chi^2/\mathrm{ndf}=\gsfRedChiSq$), reduced to $\chi^2_\mathrm{corr}=\gsfChiSqCorr$ after
the two-pass correction ($\chi^2_\mathrm{corr}/\mathrm{ndf}=\gsfRedChiSqCorr$; because this
ratio is below $1$ the covariance is left unscaled rather than shrunk,
Sec.~\ref{sec:fitting}), using \gsfNumEscales{} energy-scale
offsets. These are the values of the SIBYLL-2.3e baseline fit
(Sec.~\ref{sec:results:mixture}). That a single
smooth model reproduces direct and air-shower measurements across eleven
decades in energy, once energy scales are adjusted within their quoted
systematics, is itself a nontrivial result, given the historical tensions
between experiments in overlapping ranges. The solution is robust: replacing
the Gaussian offset penalty by a flat window leaves flux and composition
essentially unchanged, doubling the knot density does not alter the result
significantly (SM~S1), and removing individual experiments moves the
best-fit fluxes well within the quoted band.

\subsection{Spectrum and mass composition}
\label{sec:results:spectrum}
\label{sec:results:composition}

\begin{figure*}
\centering
\includegraphics[width=0.95\textwidth]{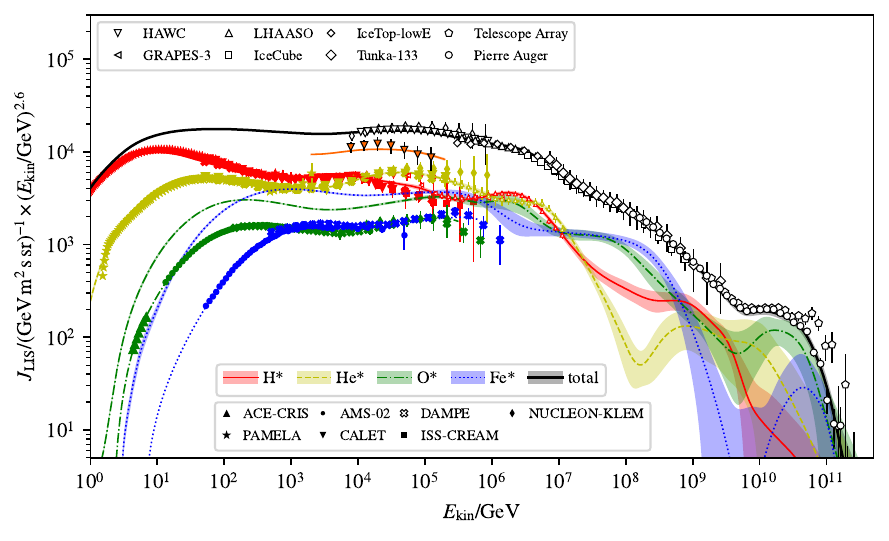}\\[6pt]
\includegraphics[width=0.95\textwidth]{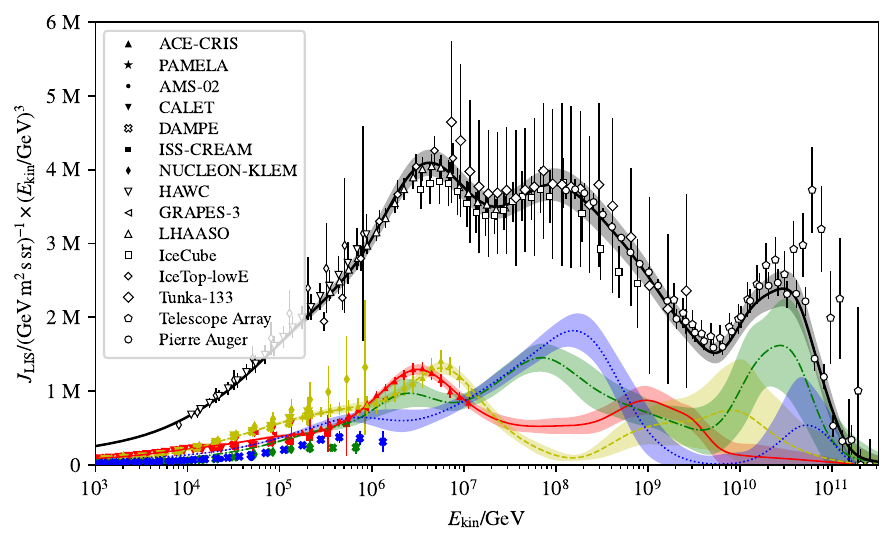}
\caption{The cosmic-ray flux from $\sim$1\gev{} to $10^{11}$\gev{}. Top:
solid
markers show direct measurements of the leading elements; open markers show
air-shower measurements of the all-particle flux and of mass groups. All data
are adjusted to the common energy scale found by the fit, and the direct data
are demodulated to the local interstellar spectrum with each experiment's
window-averaged modulation potential (SM~S2), the frame in which the model is
drawn. Lines and bands show
the \GSF{} all-particle and group fluxes with their $1\sigma$ uncertainties.
Bottom: the same fluxes, scaled by $E^{3}$ and drawn with a
linear flux axis, emphasizing the structures between the knee and the
suppression.}
\label{fig:money}
\label{fig:knee}
\end{figure*}

\begin{figure*}[p]
\centering
\includegraphics[width=0.90\textwidth]{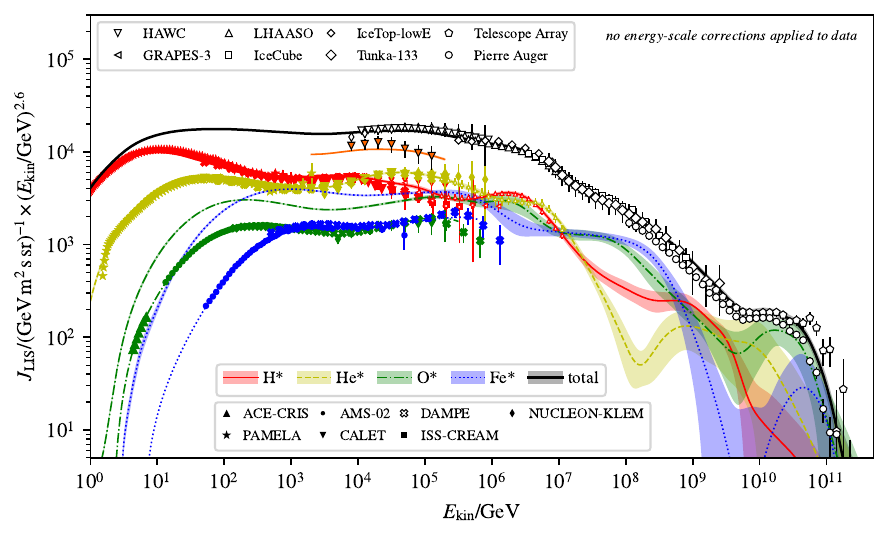}
\caption{Identical to the top panel of Fig.~\ref{fig:money}, but with the
data at their published energy scales.}
\label{fig:moneyraw}
\end{figure*}

\begin{figure*}[p]
\centering
\includegraphics[width=0.90\textwidth]{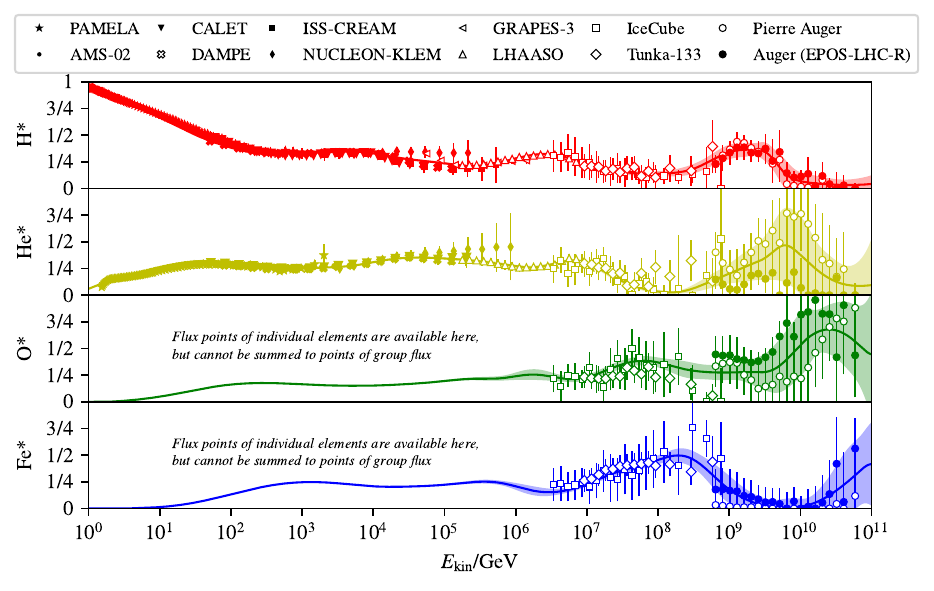}
\caption{Fractions of the four mass groups as a function of energy from
composition-sensitive air-shower measurements, compared to the \GSF{}
($\pm1\sigma$ bands). Data are adjusted to the common energy scale. The
Auger fluorescence composition is shown under both hadronic-interaction
models: open circles denote the SIBYLL-2.3e interpretation (fitted), filled
circles EPOS-LHC-R (overlay, not fitted;
Sec.~\ref{sec:results:mixture}). Curves and data are in the same frame: the
direct spectra entering the proton and helium panels are demodulated with each
experiment's window-averaged modulation potential (SM~S2).}
\label{fig:fractions}
\end{figure*}

Figure~\ref{fig:money} shows the central result of this work: the cosmic-ray
flux and its decomposition into mass groups from $\sim$1\gev{} to
$10^{11}$\gev{}, fit to the measurements of the experiments described in
Sec.~\ref{sec:data}. Figure~\ref{fig:moneyraw} shows the same comparison
with the data at their published energy scales
(Sec.~\ref{sec:results:scales}). All
established spectral features emerge from the data without being built into
the model: the hardening of the proton and helium spectra near
200--300\gv{} and their softening near 15~TV, the light-component knee
measured by LHAASO with the helium knee at 6--7~PeV, the second knee, the
ankle, the instep, and the suppression at the highest energies. The
positions of these features follow an approximately rigidity-ordered
sequence.

In the knee region the composition is anchored from below: the heavy
(O$^{*}$ and Fe$^{*}$) fluxes entering the LHAASO range are constrained by the
direct elemental DAMPE spectra, so that a large decrease of the heavy
contribution could occur only within the narrow gap between the DAMPE and
LHAASO coverages. At the lowest energies, where few data sets contribute and
their precision is already high, the fit follows the elemental data closely.
Above a few \gv{} the model band is set by the AMS-02 statistics and is
narrower than the scatter between experiments, which is dominated by
normalization differences between the calorimetric instruments and AMS-02
(CALET and DAMPE in particular); below a few \gv{} the solar-modulation
nuisance instead dominates the band and widens it beyond that scatter. This
band-coverage question is discussed in SM~S3.

\begin{figure}
\includegraphics[width=\columnwidth]{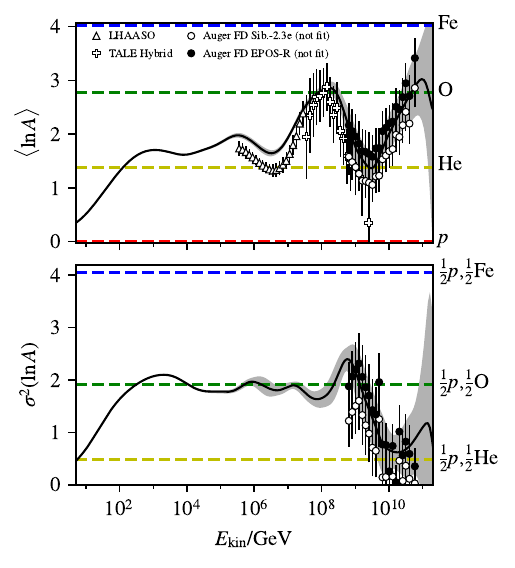}
\caption{Mean logarithmic mass (top) and its variance (bottom) as a function
of energy, computed from the \GSF{}, compared with the $\mlna$ data used in
the fit (LHAASO, TALE hybrid). Open and filled circles show the
$\mlna$ and $\sigma^2(\lna)$ derived by Auger from its
$X_\mathrm{max}$ moments under the SIBYLL-2.3e and EPOS-LHC-R
interpretations~\cite{PierreAuger:2026qbt}, adjusted to the common energy
scale by the fitted Auger offset. These moments are \emph{not} used in the
fit, since they derive from the same events as the fitted Auger mass
fractions; they are shown for comparison only.}
\label{fig:lna}
\end{figure}

Figure~\ref{fig:fractions} compares the fitted group fractions with the
composition-sensitive measurements, and Fig.~\ref{fig:lna} shows the mean
logarithmic mass and its variance. The composition grows gradually heavier
from the knee toward the ankle region and turns lighter beyond it,
consistent with the Auger and Telescope Array observations; the elongation
break observed by the TALE hybrid analysis near $10^{17.1}$\,eV is
reproduced. The remaining composition tensions, most notably between the
LHAASO flux measurements and the LHAASO $\mlna$, and the interpretation of
the $\sigma(\lna)$ band structure are collected in
Sec.~\ref{sec:discussion}.

\begin{figure}
\includegraphics[width=\columnwidth]{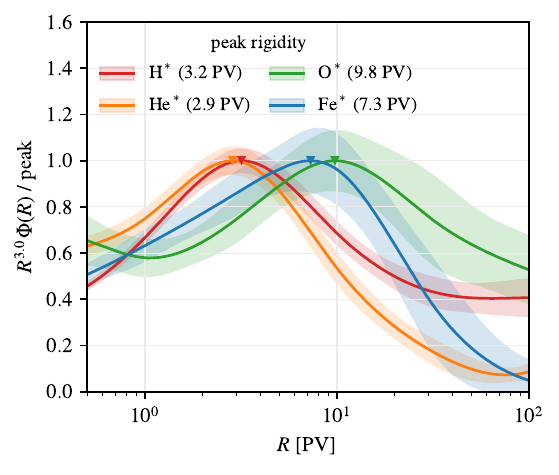}
\caption{The four \GSF{} group fluxes in the knee region as a function of
rigidity, weighted by $R^{3}$ and normalized to their maxima (triangles),
with $1\sigma$ bands.}
\label{fig:rigidity}
\end{figure}

In rigidity (Fig.~\ref{fig:rigidity}), the knees of the H$^{*}$, He$^{*}$,
and Fe$^{*}$ groups are compatible with a common cutoff rigidity, as
expected if the knee is set by a rigidity-dependent process at the sources
or in Galactic propagation, whereas the cutoff of the O$^{*}$ group is less
clearly identified. A more significant test requires less model-dependent
composition measurements in the second-knee region.

\subsection{Energy scales}
\label{sec:results:scales}
\begin{figure}
  \includegraphics[width=\columnwidth]{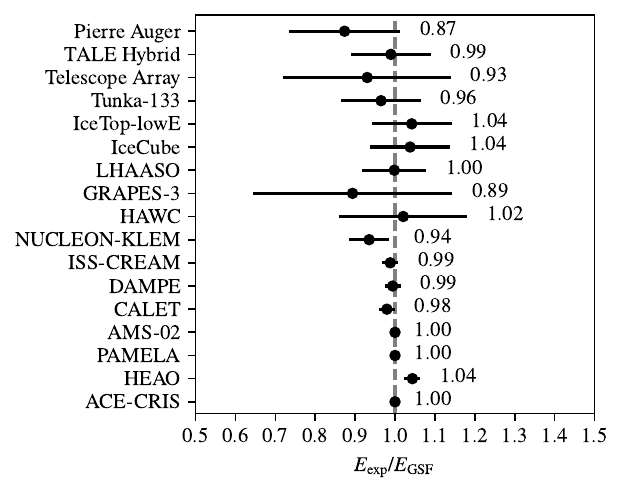}
  \caption{Fitted energy-scale factors $f_e=\tilde E/E$ per experiment. Values
  below unity indicate that an experiment's energy scale is high relative to
  the common scale of the fit. Error bars show the quoted $1\sigma$
  energy-scale systematics; the spectrometer scales (AMS-02, PAMELA, ACE-CRIS)
  are treated as exact and anchor the ensemble.}
  \label{fig:scales}
\end{figure}

Figure~\ref{fig:scales} shows the fitted energy-scale factors $f_e$. All lie
within the systematic uncertainties quoted by the experiments. The
spectrometer scales anchor the fit at the percent level; among the
air-shower experiments the fit finds, for example, $f_e\simeq \gsfEscaleLhaaso$ for
LHAASO, \gsfEscaleIceCube{} for IceCube, \gsfEscaleTunka{} for Tunka-133,
\gsfEscaleTA{} for the Telescope Array,
and \gsfEscaleAuger{} for the Pierre Auger Observatory. The offsets of the
ultra-high-energy observatories form a ladder: each experiment overlaps the
next in energy, and the spline continuity chains their relative scales down
to the direct-measurement anchor. In the transition region between direct
measurements and air-shower arrays ($\sim$0.3--10~PeV) the all-particle
normalizations are not in perfect agreement: the low-energy IceTop
spectrum sits systematically high, by up to $\sim$8\%, relative to LHAASO
and the coincident IceCube/IceTop analysis~\cite{IceCube:2019hmk} at a
common energy scale. The fit
absorbs these minor tensions through the energy-scale offsets, which
remain well within the quoted systematics ($f_e\simeq \gsfEscaleIceTopLowE$ for the
low-energy IceTop spectrum against the values above); the mass
composition is unaffected, since the tension concerns a pure all-particle
normalization. Because the composition observables
participate in the weight correction (Sec.~\ref{sec:fitting}), they pull
the air-shower scales upward by 2--3\% relative to a flux-only weighting,
and the well-known $\sim$10\% relative offset between the Telescope Array
and Auger is reduced to about \gsfEscaleTAtoAugerPct\% in the global solution. The residual scale difference appears as a
coherent $\sim$10\% flux excess of the Telescope Array spectrum over the
fitted model through the ankle decade (below $1.5\sigma$ per point), so that the all-particle
spectrum above the ankle effectively follows Auger: at the fitted scales
the Auger points scatter within $\pm$2\% of the model while Auger carries
about twice the Telescope Array's statistical weight in the common energy
range, and an order of magnitude more once its below-the-ankle spectrum,
which anchors the spline through continuity, is included. A larger
Telescope Array offset would not improve the joint fit, because the
residual is energy-dependent: it grows above $\sim 3\times10^{19}$\,eV
into the known high-energy Telescope Array--Auger discrepancy, which an
energy-independent rescaling cannot remove without over-correcting the
ankle region where the Telescope Array's weight is concentrated. The
fitted Auger offset is itself a single energy-independent factor,
anchored mainly by the below-the-ankle flux of the SD-750 array, which
shares the fluorescence energy scale of the full data set.
The cumulative size of the cross-normalization is illustrated in
Fig.~\ref{fig:moneyraw}, which shows the data of Fig.~\ref{fig:money} at
their published energy scales: the all-particle
normalizations disagree at the 10--30\% level.
The coherent downward trend of the ensemble is
discussed in Sec.~\ref{sec:discussion:scales}.

\subsection{Comparison with previous parametrizations}
\label{sec:results:models}
\begin{figure*}
\centering
\includegraphics[width=\textwidth]{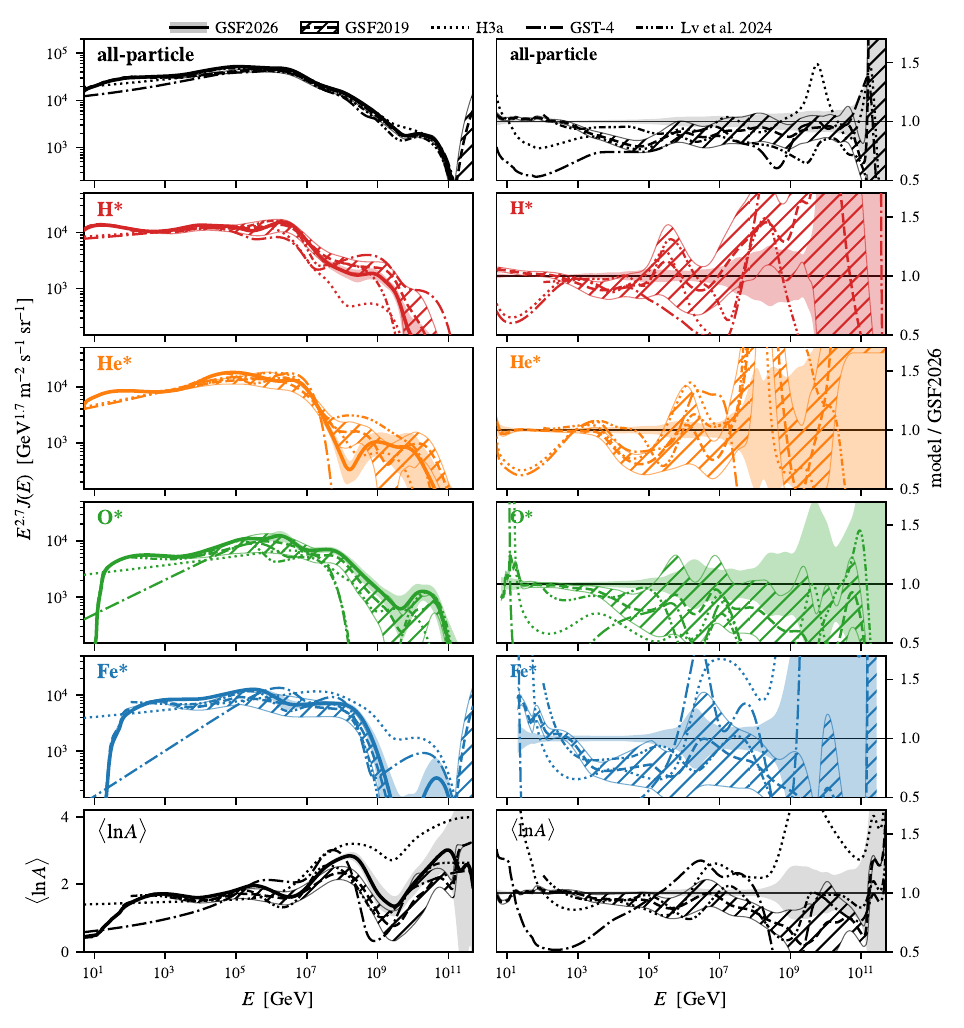}
\caption{Mass-group decomposition of the flux and the mean logarithmic mass.
The \GSF{}2026 fit (solid line, filled $\pm1\sigma$ band) is compared with
the earlier \GSF{}2019 fit (dashed line, hatched $\pm1\sigma$
band) and with the parametrizations of Gaisser
(H3a, dotted)~\cite{Gaisser:2011klf}, Gaisser--Stanev--Tilav (GST-4,
dash-dot)~\cite{Gaisser:2013bla}, and Lv et al.\ (dash-dot-dot, drawn only
above its $100$\gev{} all-particle validity floor)~\cite{Lv:2024wrs}. Rows
from top to bottom: all-particle flux, the hydrogen (H$^{\ast}$), helium
(He$^{\ast}$), oxygen
(O$^{\ast}$) and iron (Fe$^{\ast}$) groups, and $\langle\ln A\rangle$. Left
column: spectra weighted by $E^{2.7}$, with $\langle\ln A\rangle$ on a linear
scale; right column: ratio to the \GSF{}2026. Color denotes the mass group
in the \GSF{} scheme. The composition of each external model is summed into
the four \GSF{} leading groups (H$^{\ast}$ $Z=1$, He$^{\ast}$ $Z=2$, O$^{\ast}$
$Z=3$--$9$, Fe$^{\ast}$ $Z=10$--$28$); its $\langle\ln A\rangle$ uses the true
mass of each component.}
\label{fig:massgroupmodels}
\end{figure*}
\begin{figure}
\includegraphics[width=\columnwidth]{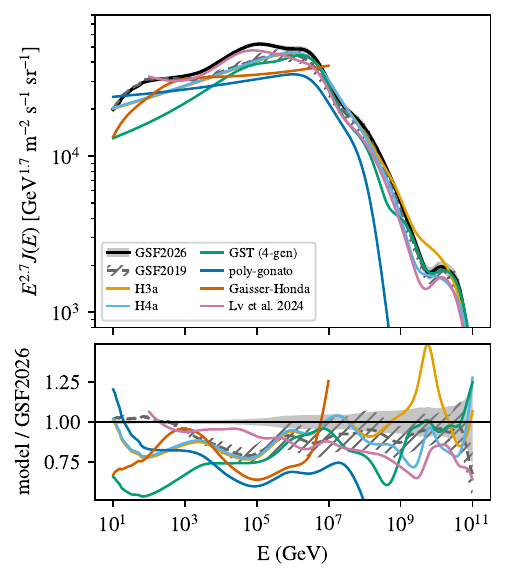}
\caption{All-particle flux of the \GSF{} (black line with $\pm1\sigma$ band)
compared with widely used broken-power-law parametrizations
\cite{Hoerandel:2002yg,Gaisser:2011klf,Gaisser:2013bla,Gaisser:2002jj} and
the LHAASO-anchored model of Lv et al.~\cite{Lv:2024wrs}.
Bottom: ratio to the \GSF{}.}
\label{fig:models}
\end{figure}

Figure~\ref{fig:models} compares the \GSF{} all-particle flux with the
broken-power-law parametrizations in common
use~\cite{Hoerandel:2002yg,Gaisser:2011klf,Gaisser:2013bla,Gaisser:2002jj}
and the recent LHAASO-anchored model of Lv et al.~\cite{Lv:2024wrs}.
Deviations of 20--50\% appear over wide energy ranges, most prominently
between $10^{4}$ and $10^{9}$\gev{}, where the new direct and knee-region
data constrain the \GSF{} to a few percent. The older parametrizations
pre-date most of these measurements, and their smooth power-law segments
average over the structures that the data now resolve; the comparison
quantifies how far the empirical knowledge of the flux has moved in the
past decade.

Figure~\ref{fig:massgroupmodels} resolves the same comparison by mass group
and extends it to the mean logarithmic mass. It also includes our earlier
\GSF{}2019 fit~\cite{Dembinski:2017zsh,Schroder:2019agg}, evaluated from its
original published parameter set, with its uncertainty band. Each
parametrization is decomposed into the four \GSF{} leading groups by summing
its elemental components. Up to the knee the models agree with the
direct-measurement groups at the 10 to 30\% level. In the
$10^{4}$--$10^{5}$\gev{} range, where the \GSF{}2019 iron group was
constrained only by sparse measurements and formed a near-flat direction of
that fit, the flux is now constrained directly by the DAMPE iron
spectrum~\cite{DAMPE:2025opn}. Above the knee the light
and heavy groups diverge, since each model places its rigidity-dependent
cutoffs at a different energy, and the spread in $\langle\ln A\rangle$
follows the differing knee compositions.

\subsection{Composition at the highest energies under the
hadronic-interaction-model mixture}
\label{sec:results:mixture}

The Auger fluorescence composition is provided in two interpretations,
obtained with the post-LHC
hadronic-interaction models SIBYLL-2.3e and EPOS-LHC-R, each with per-bin
posterior correlations. The two interpretations imply such different
compositions that averaging them within a single likelihood would yield a
blend that reflects neither. To propagate this discrete ambiguity into the
modeled uncertainty we perform two complete fits that are identical in every
respect (data selection, energy-scale treatment, weight correction, and
covariance construction) except for which model interpretation of the Auger
fractions enters, and combine them at the \emph{parameter} level as an
equal-weight two-point mixture: the amplitude vector is the mean,
$\theta=\tfrac{1}{2}(\theta_\mathrm{SIB}+\theta_\mathrm{EPOS})$, and the
covariance is
\begin{equation}
\begin{aligned}
\Sigma &= \tfrac{1}{2}\left(\Sigma_\mathrm{SIB}+\Sigma_\mathrm{EPOS}\right)
+ \tfrac{1}{4}\,\Delta\theta\,\Delta\theta^{T},\\
\Delta\theta &= \theta_\mathrm{SIB}-\theta_\mathrm{EPOS},
\end{aligned}
\label{eq:mixture}
\end{equation}
whose rank-one between-model term is fully correlated across all spline
amplitudes. The resulting band spans both single-model solutions coherently
wherever they differ and collapses to the ordinary fit covariance where they
agree. Below $\approx 2\times10^{8}$\gev{} the SIBYLL-2.3e and EPOS-LHC-R
interpretations coincide, so the two fits are indistinguishable there and the
model choice does not propagate to the knee or second-knee region. The two
component fits are of equal statistical quality,
$\chi^2_\mathrm{corr}/\mathrm{ndf}=\gsfChiSqCorr/\gsfNdof=\gsfRedChiSqCorr$ under SIBYLL-2.3e and
$\gsfChiSqCorrEpos/\gsfNdofEpos=\gsfRedChiSqCorrEpos$ under EPOS-LHC-R, so the fit quality does not discriminate
between the interpretations, and the choice of SIBYLL-2.3e as the baseline rests
on the composition-continuity argument of Sec.~\ref{sec:data}. Being an average
of two minima rather than a minimum itself, and combining fits to different
fraction data, the mixture has no $\chi^2$ of its own; the values quoted in
Sec.~\ref{sec:results:quality} are those of the SIBYLL-2.3e baseline. The
mixture is provided in the standard \GSF{} parameter-plus-covariance format
and evaluates like any single fit. Since
the underlying ambiguity is between two discrete hypotheses, the
distribution is bimodal, and the $\pm1\sigma$ band is a conservative smooth
rendering rather than a probabilistic statement about intermediate
compositions. The fitted energy scales are insensitive to the swap (Auger
$f_e$ changes by $+0.4$\%): the ultra-high-energy scale ladder is set by the
flux, not by the fractions. Figure~\ref{fig:fractions} shows the Auger
data under both interpretations.

\section{Discussion}
\label{sec:discussion}

\subsection{The energy-scale systematic}
\label{sec:discussion:scales}

The ensemble of fitted energy scales skews
only marginally low (mean offset $\approx-0.2\sigma$), with the absolute
normalization anchored at low energies by the rigidity-referenced
interstellar spectrum, the solar modulation, and the measured spectral
features. This small shift is carried in full by the published covariance and
requires no separate treatment.

\subsection{Composition in the knee region: LHAASO flux versus $\mlna$}
\label{sec:discussion:lhaaso}

One composition tension stands out. The fit reproduces the light-component
fraction implied by LHAASO's own flux measurements, with the fitted
(p+He)/all ratio matching the LHAASO fluxes at the percent level across
0.3--3~PeV. This composition, however, corresponds to a $\mlna$ that is
0.2--0.35 \emph{heavier} than the $\mlna$ published by LHAASO from the same detector
(1.63 versus 1.32 at the $10^{15.5}$\,eV dip). The resulting pulls of
$+1.7$ to $+2.8\sigma$ are robust: the $\mlna$ used here is the mean of the
three published hadronic interpretations and its uncertainty carries their
spread, which at the dip ($\pm0.07$) is a quarter of the discrepancy, so no
choice of interpretation removes it; and neither additional spline
freedom nor the removal of other data sets relieves the tension, since the
heavy component entering the LHAASO range is anchored by the direct DAMPE
spectra.

This points to a tension between the LHAASO flux and $\mlna$ analyses that
the fit, which follows the fluxes, cannot fully reconcile with the lighter
$\mlna$. Two directions could ease it: an underestimated systematic in one of
the two analyses, or a knee composition dominated by elements lighter than
the nitrogen--oxygen mass scale of the intermediate group (carbon, or even
the Be--B range), which a four-group model cannot express at the measured
light fraction. The recent LHAASO-anchored model of Lv et
al.~\cite{Lv:2024wrs} reaches the lighter $\mlna$, but does so by carrying an
iron flux well below (roughly $0.4\times$) the level now measured directly by
DAMPE~\cite{DAMPE:2025opn}, and by fitting the LHAASO all-particle flux and
$\mlna$ without the LHAASO proton and helium spectra~\cite{LHAASO:2025mlf}
that determine the light component here. The direct DAMPE iron measurement, which
post-dates the anchors of that model, thus disfavors the light-composition
route and sharpens the tension rather than removing it. We leave it as
reported; it is a concrete target for the next generation of knee-region
composition analyses.

\subsection{The dispersion of the mass distribution}
\label{sec:discussion:slna}

The variance $\sigma^2(\lna)$ shown in Fig.~\ref{fig:lna} is a pure
prediction of the fit: no dispersion measurement enters the likelihood, so
the curve and its band follow entirely from the fitted elemental fluxes and
their covariance. Since many elemental combinations produce the same
dispersion, the equal two-component mixtures indicated by dashed lines are
visual references only. The composition remains mixed, with
$\sigma^2(\lna)\approx 2$, over roughly seven decades in energy and turns
nearly pure only at the highest energies, where the unfitted Auger moments
reach values consistent with zero, reflecting their narrow
$X_\mathrm{max}$ distributions~\cite{PierreAuger:2026qbt}. The $\mlna$
panel also illustrates the hadronic-model policy of Sec.~\ref{sec:data}:
since the TALE $\mlna$ is interpreted with QGSJET-II-04, fitting the Auger
fractions under EPOS-LHC-R alone would imprint an artificial step in
$\mlna$ between the two energy ranges, which the SIBYLL-2.3e baseline
avoids.

\section{The nucleon flux}
\label{sec:nucleon}
\begin{figure}
\includegraphics[width=\columnwidth]{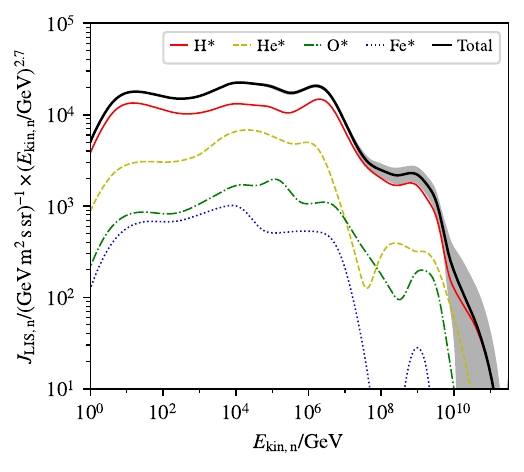}
\caption{Nucleon flux versus energy per nucleon from the \GSF{} (thick
line, $\pm1\sigma$ band) with the contributions of the four mass groups.
Dots mark the spline knots mapped to equivalent nucleon energy.}
\label{fig:nucleon}
\end{figure}

Atmospheric lepton fluxes are calculated from the spectrum of nucleons
arriving at the atmosphere, conveniently expressed per energy-per-nucleon
$E_N=E/A$: a nucleus $(Z,A)$ contributes $Z$ protons and $A-Z$ neutrons, so
that
\begin{equation}
J_N(E_N) = \sum_i \bigl[ Z_i + (A_i-Z_i) \bigr]\, A_i\,
J_i\bigl(A_i E_N\bigr),
\label{eq:nucleon}
\end{equation}
summed over all elements $i$, with the proton and neutron parts entering
atmospheric cascades separately. The nucleon flux is not directly observed;
by Eq.~(\ref{eq:nucleon}) it is a derived quantity of the
\GSF{}, with uncertainties propagated through the full covariance. Figure~\ref{fig:nucleon} shows the result. The nucleon
flux is dominated by the proton and helium contributions, and since the
proton spectrum is now covered by data from GeV to EeV energies, its
uncertainty is correspondingly small; the sub-leading abundances play
essentially no role. The precision of the recent direct data imprints
sharply resolved features (the 200--300\gv{} hardening, the 15~TV
softening, the light-component knee) that translate directly into
structure expected in atmospheric neutrino and muon spectra.

\begin{figure}
\includegraphics[width=\columnwidth]{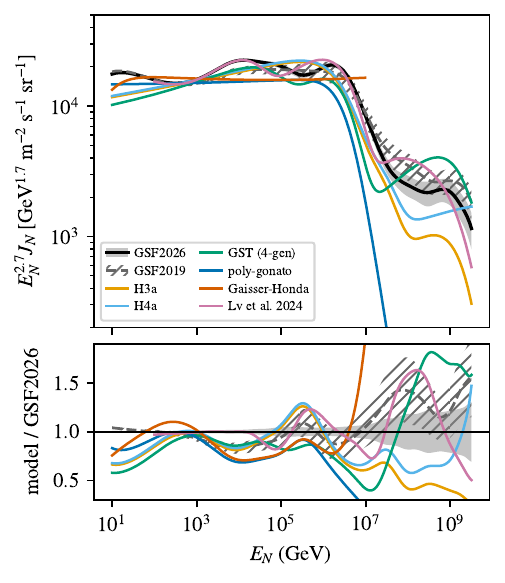}
\caption{Nucleon flux of the \GSF{} compared with the parametrizations of
Refs.~\cite{Hoerandel:2002yg,Gaisser:2011klf,Gaisser:2013bla,Gaisser:2002jj}
and the model of Lv et al.~\cite{Lv:2024wrs}.
Bottom: ratio to the \GSF{}; the gray band shows the \GSF{} $\pm1\sigma$
uncertainty.}
\label{fig:nmodels}
\end{figure}

Figure~\ref{fig:nmodels} compares the \GSF{} nucleon flux with the
parametrizations most widely used in atmospheric lepton calculations and
with the model of Lv et al. The
differences reach 20--50\% between $10^{4}$ and $10^{9}$\gev{}, exceeding
the \GSF{} uncertainty band several times, and their sources can be traced to specific
measurements: the AMS-02/DAMPE structures below 100~TV and the LHAASO
light-component knee at PeV energies. Predictions of atmospheric neutrino
and muon fluxes built on the older parametrizations inherit these deviations,
which may lie outside the typical uncertainty attributed to the primary flux in
standard estimates.

\begin{figure}
\includegraphics[width=\columnwidth]{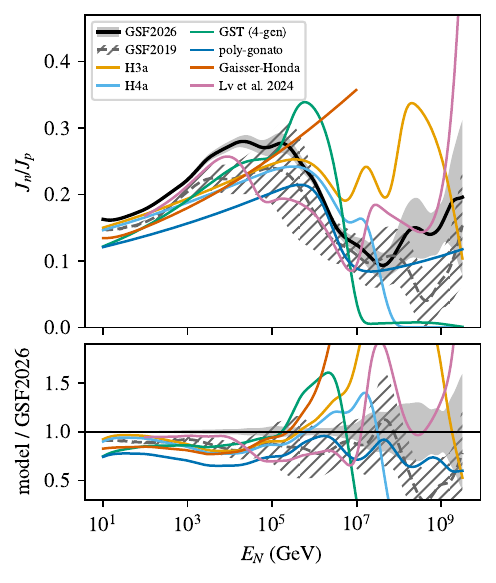}
\caption{Neutron-to-proton ratio $J_n/J_p$ of the nucleon flux from the
\GSF{} ($\pm1\sigma$ band, propagated including the proton--neutron
covariance) and its 2019 predecessor, compared with the models of
Fig.~\ref{fig:nmodels}~\cite{Hoerandel:2002yg,Gaisser:2011klf,Gaisser:2013bla,Gaisser:2002jj,Lv:2024wrs}.
Bottom: ratio to the \GSF{}.}
\label{fig:npratio}
\end{figure}

A second input to atmospheric-lepton calculations is the neutron-to-proton
ratio of the nucleon flux (Fig.~\ref{fig:npratio}), which sets the charge
asymmetry of the secondaries, most directly the atmospheric muon charge
ratio~\cite{Gaisser:2011klf}. Since neutrons arrive only bound in nuclei,
$J_n/J_p$ is a direct image of the mass composition. Below the knee the
\GSF{} composition is heavier than assumed in the earlier parametrizations,
most prominently around 10~TeV per nucleon, where the hard helium spectrum
of the modern direct data meets the 15~TV proton softening; the neutron
fraction is correspondingly larger. Deuterium, which the \GSF{} now
includes as a separate species, is a significant contributor to the change
with respect to the 2019 version: it raises $J_n/J_p$ by about 0.01 at all
energies and dominates the difference below 1~TeV per nucleon. Muon charge
ratios computed from the \GSF{} nucleon fluxes are therefore expected to
come out slightly lower than those built on the older inputs.

For applications that constrain or propagate these uncertainties in their
own fits, the full parameter covariance is impractical. We therefore
provide a compact nuisance-parameter representation, which deforms the
proton and neutron fluxes relative to the central model,
\begin{equation}
J_s(E;\boldsymbol{\theta}) = J_{s,\mathrm{central}}(E)\,
\bigl[ 1 + \textstyle\sum_k H_k(E)\,\theta_{s,k} \bigr],
\label{eq:reduced}
\end{equation}
where each component $\theta_{s,k}$ is the relative flux deviation at one
of twelve fixed pivot energies and the cardinal functions $H_k$ interpolate
smoothly between them. The 24 components carry the exact covariance of the
full fit at the pivot energies, to be imposed as a Gaussian penalty. The
construction serves the same role as the principal-component
parametrization of the \GSF{} uncertainty used in
daemonflux~\cite{Yanez:2023lsy}, with directly interpretable parameters in
place of eigenmode amplitudes. Between the pivots the reduced uncertainty matches the full one
within a factor of 1.22 at worst over $1$--$10^{9}$\gev{} per nucleon,
which covers atmospheric-lepton applications; the construction and its
validation are described in SM~S4.

\section{Summary and outlook}
\label{sec:summary}

This work delivers a data-driven determination of the cosmic-ray flux and
mass composition from $\sim$1\gev{} to $10^{11}$\gev{}, built from the most
precise direct and air-shower measurements available and deliberately
agnostic about the underlying astrophysics. With per-experiment energy
scales cross-calibrated within their quoted systematics, a single smooth
model describes about one thousand data points with
$\chi^2/\mathrm{ndf}\approx \gsfRedChiSqShort$, or $\gsfRedChiSqCorr$ after the localized
disagreements are de-weighted by the PDG scale-factor prescription
(Sec.~\ref{sec:fitting}): once the energy scales are aligned, the global
body of cosmic-ray data forms a consistent picture, to a degree that was
not demonstrable a decade ago. The
progress of direct measurements, the LHAASO knee-region data, the TALE
coverage of the second knee, and the Auger fluorescence composition shrink
the uncertainties substantially with respect to the 2017
fit~\cite{Dembinski:2017zsh}, without qualitatively changing its picture, which speaks for the
stability of the method. A small number of data sets was not considered
for the \GSF{}2026 parametrization,
for documented inconsistencies, and the principal remaining tension, between
the LHAASO flux and $\mlna$ analyses, is stated as a result rather than
absorbed. The approximate alignment of the group knees in rigidity
resembles a Peters cycle~\cite{Peters:1961mxb}, but a decisive test
requires more precise heavy-group spectra in the second-knee region.

Future measurements will sharpen this picture further. At air-shower
energies, the most valuable additions would be a full four-group composition
result from the TALE hybrid analysis with its covariance, an updated IceCube
composition analysis interpreted with more recent post-LHC hadronic models,
direct heavy-element measurements extending into the PeV range covered by
LHAASO, a closer consensus among hadronic-interaction models at the highest
energies, and improved Telescope Array results, with quantified systematics,
mass-group covariances, and modern hadronic models for the composition
inference, providing northern-sky coverage to complement Auger. In the
direct-measurement range, identifying the sources of the present tensions
between PAMELA and CALET and AMS-02, and between DAMPE and LHAASO at their
energy crossover near 100~TeV, would close the remaining gaps toward an
unambiguous, high-precision view of the cosmic-ray spectrum and its chemical
composition.

The most immediate application is to
atmospheric lepton fluxes: the nucleon flux derived here differs from the
parametrizations underlying current neutrino-flux predictions by far more
than their nominal primary-flux uncertainties, and carries sharply resolved
features that older models lack. Updating these predictions, and their
error budgets, is the natural next step.

\paragraph*{Software availability.}
The \GSF{} model, comprising the fitted parameter sets, their covariances,
and the reduced representation, is distributed as the open-source Python
package \texttt{globalsplinefit}, available on
PyPI and developed at
\url{https://github.com/gsf-project/globalsplinefit}. Its documentation
includes the GSF Explorer, a web interface to the model, and interactive
examples that run in the browser. The fitting code used to produce the parameter sets is
not public. Requests for access that are specific and scientifically motivated
will be considered by the authors on a case-by-case basis.

\begin{acknowledgments}
We thank our colleagues from the IceCube Neutrino and Pierre Auger
Observatories, Alfredo Ferrari, Javier G.~Gonzalez, and Armando di Matteo
for valuable comments. We dedicate this work to the memory of Thomas
K.~Gaisser, who contributed to the initial vision of this model.
AF and KF acknowledge support from Academia Sinica grants AS-GCS-113-M04, AS-CDA-115-M01,
and the National Science and Technology Council grant 113-2112-M-001-060-MY3. This work was
supported in part by the Academia Sinica Grid Computing Center (AS-GCC) Grand Challenge Grant No.~AS-GCS-113-M04.
\end{acknowledgments}

\bibliography{gsf2026}

\clearpage
\onecolumngrid
\begin{center}
{\large\bfseries Supplemental Material\par}\medskip
{\bfseries Global Spline Fit: A unified data-driven view
of the cosmic-ray spectrum and mass composition from GeV to the highest
energies\par}
\end{center}
\bigskip
\twocolumngrid
\setcounter{section}{0}
\setcounter{figure}{0}
\setcounter{table}{0}
\setcounter{equation}{0}
\renewcommand{\thesection}{S\arabic{section}}
\renewcommand{\thetable}{S\arabic{table}}
\renewcommand{\thefigure}{S\arabic{figure}}
\renewcommand{\theequation}{S\arabic{equation}}
\renewcommand{\theHsection}{S\arabic{section}}
\renewcommand{\theHtable}{S\arabic{table}}
\renewcommand{\theHfigure}{S\arabic{figure}}
\renewcommand{\theHequation}{S\arabic{equation}}

\section{Units, kinematics, and the spline basis}
\label{sm:knots}

\subsection{Kinematics}

For a species with charge $Z$ and mass number $A$ (average nucleon mass
$m_A$), total energy and rigidity are related by
\begin{align}
E &= \sqrt{(ZeR)^2 + (A m_A c^2)^2},\nonumber\\
Ze\,\frac{\mathrm{d}R}{\mathrm{d}E} &=
\left[1+\left(\frac{A m_A c^2}{ZeR}\right)^2\right]^{1/2},
\label{eq:kinematics}
\end{align}
and fluxes transform as $J(E) = J(R)\,\mathrm{d}R/\mathrm{d}E$. The
all-particle flux per energy interval is
\begin{equation}
J(E) = \sum_L \sum_{j\in L} w_{Lj}\, J_L\bigl(R_j(E)\bigr)\,
\Bigl(\frac{\mathrm{d}R}{\mathrm{d}E}\Bigr)_{j},
\label{eq:sm-total}
\end{equation}
where $L$ runs over the four leading elements, $j$ over the members of each
group, and $w_{Lj}$ are the sub-leading abundance weights. Below the top of
its measured range each sub-leading element carries its own short spline,
fit directly to the elemental data (HEAO-3~\cite{HEAO_data} for beryllium
through nickel, AMS-01~\cite{AMS01_data_Li} for lithium, ACE-CRIS at the
lowest rigidities, and the recent direct measurements including the CALET
sub-iron spectra~\cite{CALET:2025dgy}); above it, $w_{Lj}$ is the
member-to-leader ratio at the matching point, extrapolated with a fitted
power-law index as described in Sec.~\ref{sm:subleading}.

\subsection{Spline basis and knot placement}

The flux of each leading element is expanded in clamped cubic basis splines
(B-splines)~\cite{DeBoor_splines}. A B-spline $b_{k,n}(x)$ of order $n$ on a
knot grid $\{x_k\}$ is defined by the recursion
\begin{align}
b_{k,n}(x) &= \frac{x-x_k}{x_{k+n}-x_k}\, b_{k,n-1}(x)\nonumber\\
&\quad + \frac{x_{k+n+1}-x}{x_{k+n+1}-x_{k+1}}\, b_{k+1,n-1}(x),\nonumber\\
b_{k,0}(x) &= \begin{cases} 1 & x_k \le x < x_{k+1}\\ 0 &
\text{otherwise,}\end{cases}
\label{eq:bspline}
\end{align}
with $n=3$ and $x=\ln(R/\mathrm{GV})$ throughout. The boundary knots are
repeated (clamped), giving $K+2$ basis functions for $K$ knots; the first and
last amplitudes are fixed to zero, which we found convenient and without
effect on the fit in the data-covered range.

\begin{table*}
\caption{Knot positions of the leading-element splines and of the deuteron
spline in the adopted \GSF{} fit, in $\log_{10}(R/\mathrm{GV})$. Sub-leading
elements (Li--Ni) carry short spline grids covering their measured ranges
only. The air-shower range grid (9.0, 9.25, 9.5, 9.75, 10.0, 10.5, 11.5) is
shared by all four leaders. Dots abbreviate a uniform continuation at the
step set by the preceding pair.}
\label{tab:knots}
\begin{ruledtabular}
\begin{tabular}{lcp{0.78\textwidth}}
Spline & $n_\mathrm{knots}$ & \raggedright Knots $\log_{10}(R/\mathrm{GV})$ \tabularnewline
\colrule
p & 26 & 0.00 0.30 0.50 0.70 1.00 1.50 2.00 3.00 3.50 $\dots$ 9.00 9.25 $\dots$ 10.00 10.50 11.50 \\
D & 6 & 0.26 0.48 0.70 0.95 1.15 1.34 \\
He & 25 & 0.00 0.30 0.50 0.70 1.00 1.50 2.00 3.00 3.50 4.00 5.00 5.50 $\dots$ 9.00 9.25 $\dots$ 10.00 10.50 11.50 \\
O & 26 & 0.00 0.18 0.30 0.50 0.70 1.00 1.50 2.00 3.00 4.00 4.50 $\dots$ 9.00 9.25 $\dots$ 10.00 10.50 11.50 \\
Fe & 22 & 0.00 1.00 1.50 2.00 3.00 4.00 4.50 $\dots$ 9.00 9.25 $\dots$ 10.00 10.50 11.50 \\
\end{tabular}
\end{ruledtabular}
\end{table*}

\begin{figure}
\includegraphics[width=\columnwidth]{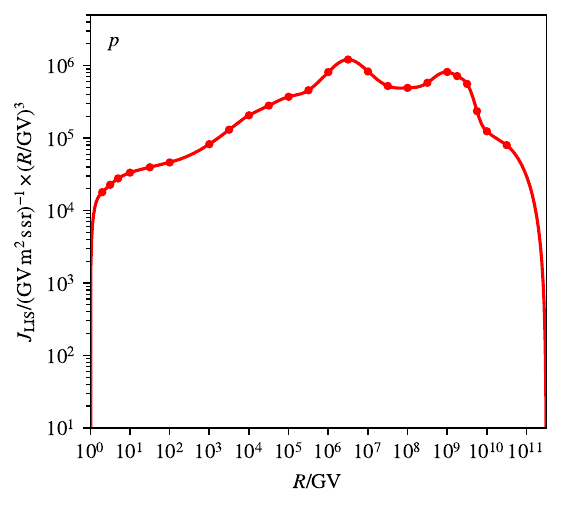}
\caption{Proton-leader spline: the fitted flux (line), the knot positions,
and the weighted B-spline basis functions that compose it.}
\label{fig:splinebasis}
\end{figure}

The knots (Table~\ref{tab:knots}) are placed approximately uniformly in
$\log_{10}(R/\mathrm{GV})$, with higher density where precise direct data
resolve spectral structure and around rapidly varying features such as the
knee, and coarser spacing across the air-shower range. Although today's data coverage would tolerate a denser grid over most
of the energy range, the composition information in the second-knee region
does not, and we deliberately keep the non-uniform grid. Uniformity and
shift tests were performed for the 2017 fit~\cite{Dembinski:2017zsh} and
repeated for the \GSF{}-2025 update~\cite{Fujisue:2025wnp}; doubling the knot
density does not change the fitted fluxes significantly and, where it changes
anything, slightly narrows the uncertainty bands, the opposite of what a
covering uncertainty should do; the coarser grid is therefore retained.
Figure~\ref{fig:splinebasis} shows the proton flux together with its weighted
basis functions.

As additional cross checks, several modifications of the ultra-high-energy
grid were tested against the instep region near $10^{19.2}$\,eV; none
improved the fit without either over-fitting the data-free proton tail or
artificially tightening the $\sigma(\lna)$ band, so the standard grid is
kept and the residual instep tension is reported in the main text. A
penalized-spline (P-spline) alternative with data-driven knot placement was
also explored: it reproduces the fit where data are dense, but it adds a
free penalty-scale parameter that cannot be reliably constrained where the
data have gaps (the composition between the second knee and the Auger
range, or the heavy components near the knee), and its smoothing-optimal
uncertainty bands are narrower than the experiment spread; it was therefore
not adopted.

\subsection{Sub-leading elements and their high-energy extrapolation}
\label{sm:subleading}

Below the top knot $R_{\mathrm{max},j}$ of its spline, each
sub-leading element $j$ interpolates the direct elemental data; the top
knot of every sub-leading spline is placed at that element's last measured
point, so the fitted spline never extends into a data-free gap and the
extrapolation takes over exactly at the end of the data, continuous with
the spline by construction. Above it,
previous \GSF{} releases~\cite{Dembinski:2017zsh} froze the
member-to-leader flux ratio at its matching-point value,
$w_{Lj} = J_j(R_{\mathrm{max},j})/J_L(R_{\mathrm{max},j})$. The measured
ratios, however, retain a clear energy dependence at the top of the
measured range, most prominently the decline of the secondary species
seen by AMS-02~\cite{AMS:2018tbl} and, for the sub-iron elements, by
CALET~\cite{CALET:2025dgy}. In this release the extrapolation therefore
carries a power-law tilt that saturates to a constant at
$R_\mathrm{sat} = 5$\,PV,
\begin{equation}
J_j(R) = w_{Lj}
\left(\frac{\min(R, R_\mathrm{sat})}{R_{\mathrm{max},j}}\right)^{\!s_j}
J_L(R),\,\, R > R_{\mathrm{max},j}.
\label{eq:subtilt}
\end{equation}
The index $s_j$ and the normalization $w_{Lj}$ are both anchored to the
data. Over the last decade of element $j$'s direct data a power-law line is
fit to the measured member-to-leader ratio (corrected for the fitted
energy-scale offsets and the solar demodulation, weighted with the inverse
squared relative errors of the points); $s_j$ is its slope and
$\bar{w}_{Lj}$ its value at $R_{\mathrm{max},j}$. This is the analogue of
the power-law fits to the last measured bins in Ref.~\cite{CALET:2025dgy},
applied uniformly to all 25 sub-leading species, and the resulting indices
are consistent with the officially published elemental
ratios~\cite{AMS:2018tbl,CALET:2025dgy}. The spline-endpoint ratio
$w_{Lj} = J_j(R_{\mathrm{max},j})/J_L(R_{\mathrm{max},j})$ that fixes the
extrapolation norm is tied to $\bar{w}_{Lj}$ by a Gaussian penalty whose
width is the log prediction error of that fit, so the endpoint follows the
error-weighted data trend rather than a single, possibly fluctuating, last
measured point; across the 25 species the fitted endpoints agree with
$\bar{w}_{Lj}$ to within one standard deviation of the fit (largest offset
$\approx0.8\sigma$). The saturation rigidity, placed at the proton knee, is
motivated by the common galactic origin and the assumed similarity of
transport effects above it; it also caps the lever arm of the tilt during
the first fit pass, while $s_j$ and the penalty are still free. After that
pass both are held fixed through the final fit and the covariance
evaluation; because the penalty couples each sub-leading top-knot amplitude
to its group leader, the covariance carries that coupling, and the
uncertainty band of a sub-leading element otherwise inherits the relative
uncertainty of its group leader, the convention of the original \GSF{}
release~\cite{Dembinski:2017zsh}.

\begin{figure*}
\centering
\includegraphics[width=0.9\textwidth]{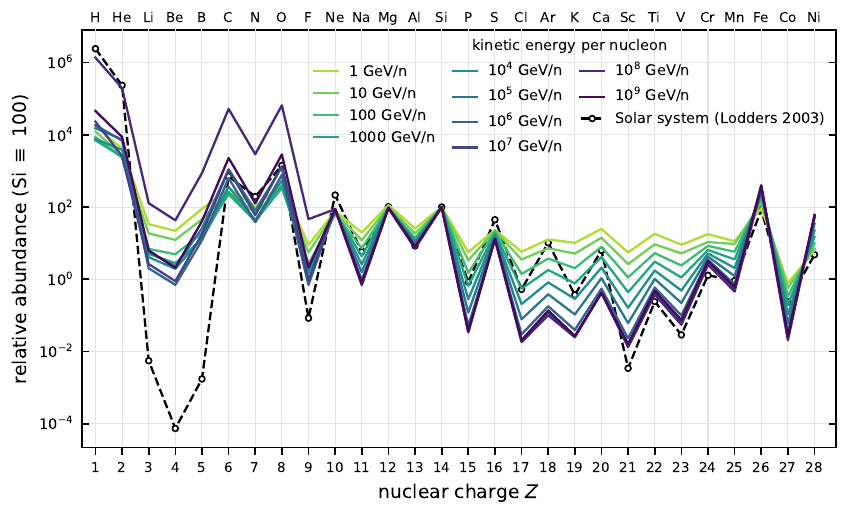}
\caption{Elemental abundances at fixed kinetic energy per nucleon from the
\GSF{} (colored lines, 1--$10^{9}$\gev/n), normalized to
$\mathrm{Si}=100$, compared with the solar-system abundances of
Ref.~\cite{Lodders:2003vvq} (dashed, open circles). With the
constant-ratio extrapolation of previous releases all curves above
$\sim$1\,TeV/n would coincide; the power-law tilts of
Eq.~\eqref{eq:subtilt} drive the secondary valleys toward the
(source-like) solar pattern as the energy increases, until the pattern
freezes at the saturation rigidity $R_\mathrm{sat}=5$\,PV.}
\label{fig:abund}
\end{figure*}

\begin{figure*}[p]
\includegraphics[width=\textwidth]{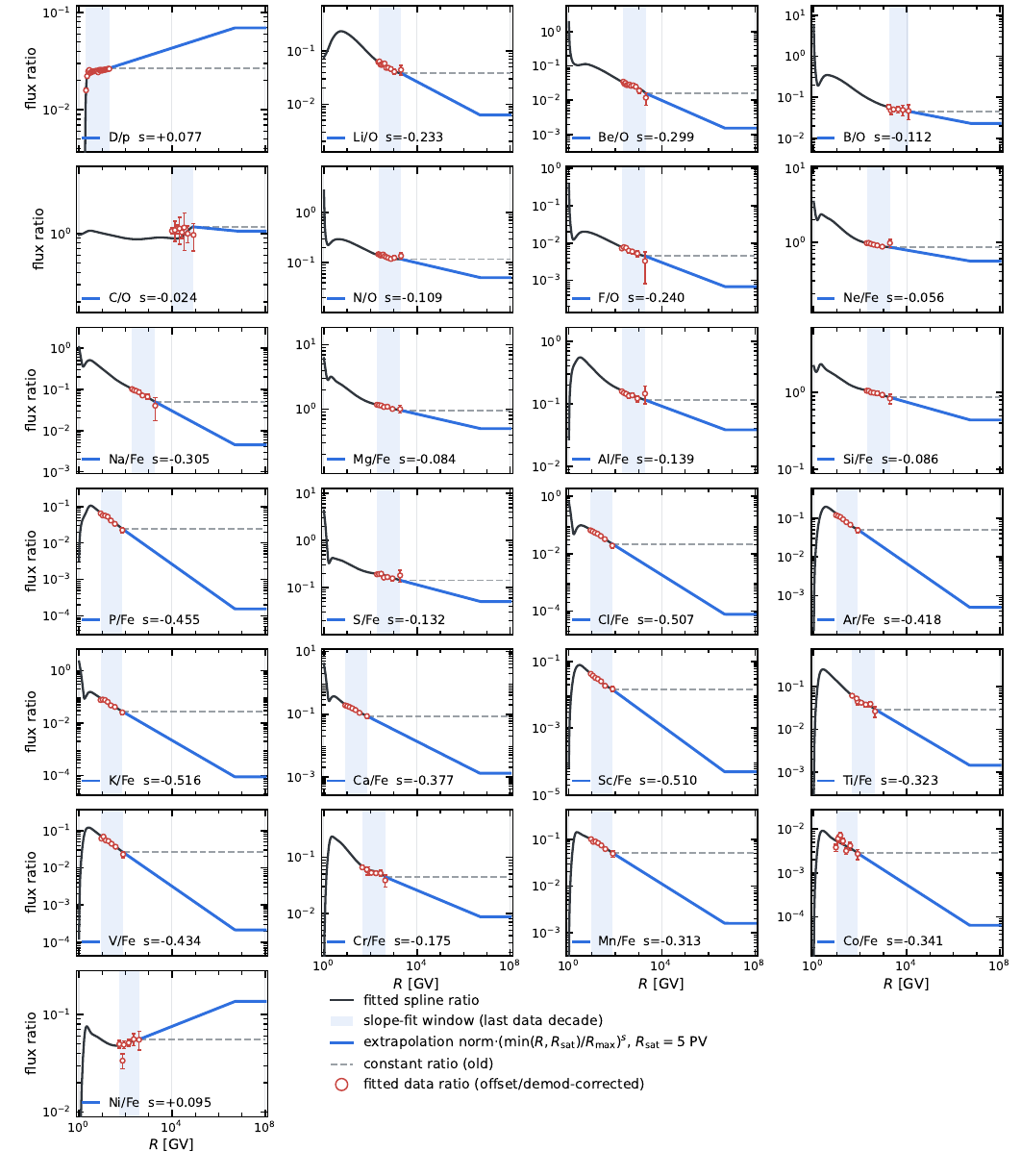}
\caption{Sub-leading--to--leader flux ratios of all 25 sub-leading species
(black: fitted spline ratio over the measured range; shaded: the
slope-fit window, the last decade of the element's direct data below the
matching point $R_{\mathrm{max},j}$; red open circles: the
offset- and demodulation-corrected member/leader data ratios that enter the
likelihood, with the weights used in the window fit). The spline endpoint is
tied by the data-anchored penalty to the error-weighted trend of the window,
so it follows the data rather than the last measured point. Beyond
$R_{\mathrm{max},j}$ the blue line shows the adopted power-law
extrapolation of Eq.~\eqref{eq:subtilt} with the fitted index $s_j$ (quoted
per panel), launched continuously from the spline endpoint and saturating
to a constant ratio at $R_\mathrm{sat}=5$\,PV; the gray dashed line is the
constant-ratio extrapolation of previous \GSF{} releases.}
\label{fig:subratios}
\end{figure*}

The impact on the fit itself is small: the air-shower observables
constrain the mass-group totals and are not sensitive to the intra-group
decomposition, so the tilt is visible only in minor shifts of $\mlna$
(at most $0.13$ between $10^{6}$ and $10^{9}$\gev) at a modestly improved
total $\chi^2$ and unchanged energy-scale offsets. The extrapolated
elemental fluxes, in contrast, change qualitatively:
Figure~\ref{fig:abund} shows how the secondary valleys of the elemental
abundance pattern continue to deepen with energy toward a source-like
composition until the pattern freezes at the saturation rigidity, and
Fig.~\ref{fig:subratios} shows the fitted ratios, slope-fit windows, and
the old and new extrapolations for all 25 sub-leading species.

\section{Likelihood definitions and uncertainty propagation}
\label{sm:likelihood}

The residual definitions below use the kinematic conventions of Sec.~\ref{sm:knots}.

\subsection{Residual definitions}

Data sets contribute to the objective through per-point residuals $r_i$,
the difference between a measured value and the corresponding model
prediction, defined in four forms below. Throughout, $i$ labels one published
data point of experiment $e$, that is an abscissa together with the value
measured there: $(\tilde R_i,\tilde J_i)$ for an elemental flux in rigidity,
$(\tilde E_i,\tilde J^\mathrm{grp}_i)$ for a group flux in energy, and
likewise $(\tilde E_i,\tilde f_i)$ and $(\tilde E_i,\widetilde{\mlna}_i)$ for a
group fraction and a mass moment. Tildes denote measured values as reported on
the experiment's native energy scale, and
$f_e = 1+\sigma_{E,e} z_e$ is the energy-scale factor of experiment $e$
(Sec.~II.B of the main text); dividing the abscissa by $f_e$ moves the point
onto the common scale, and the prefactor $1/f_e$ carries the corresponding
change of the flux normalization.

\emph{Elemental fluxes in rigidity} (direct measurements). For a point of
element $L$ (leading or sub-leading),
\begin{equation}
r_i \;=\; \tilde J_i - \tfrac{1}{f_e}\, J_L\!\bigl(\tilde R_i/f_e\bigr).
\end{equation}

\emph{Group fluxes in energy} (air showers). For a mass-group set
$G$,
\begin{equation}
r_i \;=\; \tilde J^\mathrm{grp}_i - \tfrac{1}{f_e} \sum_{L\in G} J^\mathrm{grp}_L
\bigl(\tilde E_i/f_e\bigr),
\end{equation}
with $J^\mathrm{grp}_L$ the group flux built from Eq.~(\ref{eq:sm-total})
restricted to the members of $L$.

\emph{Group fractions}. Fractions are invariant under a common energy-scale
shift of numerator and denominator,
\begin{equation}
r_i \;=\; \tilde f_i - \frac{\sum_{L\in G} J^\mathrm{grp}_L(\tilde
E_i/f_e)}{\sum_{L'} J^\mathrm{grp}_{L'}(\tilde E_i/f_e)},
\end{equation}
where $L'$ runs over all mass groups.

\emph{Mean logarithmic mass.} For $\mlna$ (and analogously its variance),
\begin{equation}
r_i \;=\; \widetilde{\mlna}_i - \mlna\!\bigl(\tilde E_i/f_e\bigr),
\end{equation}
with the model moments computed from Eq.~(2) of the main text.

The residuals of a data set are collected into its $\chi^2$ contribution
$r^{T} V^{-1} r$. Unless the experiment publishes covariance information,
the covariance matrix $V$ is built from the quoted statistical and
systematic uncertainties,
$V_{ij} = (\sigma_{\mathrm{stat},i}^2 + \sigma_{\mathrm{sys},i}^2)\,
\delta_{ij} +
\rho\,\sigma_{\mathrm{sys},i}\,\sigma_{\mathrm{sys},j}\,(1-\delta_{ij})$
with $\rho=0.5$, i.e.\ the systematic uncertainties are treated as
half-correlated across the points of the set. Where published, covariance
information replaces this default: the IceCube mass-fraction correlation
matrices (applied per energy bin across the four
fractions)~\cite{IceCube:2019hmk}, the posterior correlations of the Auger
fluorescence fractions~\cite{PierreAuger:2026qbt}, and the structural
anti-correlation
$\mathrm{Cov}(\mathrm{He},\mathrm{p}) = -\sigma^2_{\mathrm{p,stat}}$ between
the LHAASO proton and helium spectra, which derive from a common light
event sample~\cite{LHAASO:2025mlf}. The LHAASO analysis itself uses a
cosmic-ray composition model (seeded by an earlier \GSF{} release) to
subtract the heavy-element contamination of that light sample; the
published proton and helium fluxes are anchored to the measured events,
and the residual model dependence is confined to the subtracted heavy
contamination. Since these points enter our fit only through the proton
and helium components, while the heavy groups remain constrained by the
other air-shower data, the composition assumption of that analysis does
not feed back into the quantities it could bias.

\subsection{Solar modulation}
\label{sm:solarmod}

Each low-rigidity data set $e$ is compared to the local interstellar
spectrum modulated by the force-field approximation with potential
$\phi(t)$ and averaged over the experiment's observation window,
\begin{equation}
\langle J \rangle (R) \;=\; \frac{1}{N}\sum_{t\in\mathrm{window}}
J_\mathrm{FF}\bigl(R;\,\phi_t + \delta\phi_0 + \delta\phi_e\bigr),
\label{eq:smsolarmod}
\end{equation}
with $N$ the number of monthly steps in the window, using the monthly
Ghelfi--Maurin--Derome neutron-monitor series~\cite{Ghelfi:2016pcv}. The
window average matters for instruments whose exposures span varying solar
activity; for AMS-02, whose data releases combine different windows per
element, a single-epoch treatment biases the low-rigidity proton spectrum
by up to $3\%$ at $1$\gv{}, and the heavier elements by $1$--$2\%$ over their
lowest rigidity bins.

The shifts $\delta\phi_0$ and $\delta\phi_e$ propagate the uncertainty of
the reconstructed potentials into the fit: the common mode $\delta\phi_0$
(prior $\sigma_0=26$\,MV) carries the part correlated across experiments,
and the per-experiment shifts $\delta\phi_e$ move each observation window
coherently, with priors
$\sigma_e = \sigma_\mathrm{month}/\sqrt{\max(N_\mathrm{yr},1)}$ and
$\sigma_\mathrm{month}=30$\,MV, sized from the uncertainty budgets quoted
for the reconstructions~\cite{Ghelfi:2016pcv,Usoskin:2017cli}. Both enter
the objective [Eq.~(6) of the main text] as unit-Gaussian penalties,
fitted simultaneously with the amplitudes and energy-scale offsets, and
are listed in Table~\ref{tab:treatment} (Sec.~\ref{sm:datatable}). The
shifts are not degenerate with the energy-scale offsets: a potential
shift is a curved signature confined to the lowest rigidities, whereas an
offset rescales the full spectrum. The fitted common mode is small, the
largest per-experiment shifts occur for the two solar-minimum snapshot
windows (ACE-CRIS and HEAO-3) in physically sensible directions, and the
main effect on the published model is a widening of the low-rigidity
uncertainty band (for protons at $2$\gv{}, from $1.2$\% to $5.1$\%);
above $\sim 20$\gv{} the band reverts to the nuisance-free fit.

\begin{figure}
\includegraphics[width=\columnwidth]{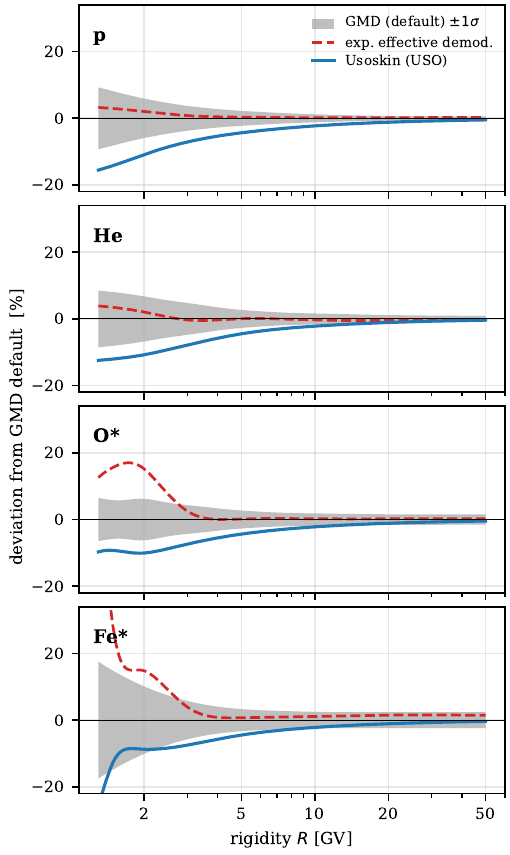}
\caption{Impact of the solar-modulation potential on the fitted local
interstellar spectra of the four mass groups, relative to the default
(Ghelfi--Maurin--Derome, parameter set \texttt{2026.1}). The gray band is the
$1\sigma$ uncertainty of the default. Blue: the Usoskin~2017
potential~\cite{Usoskin:2017cli} with the same observation-window average and
$\delta\phi$ nuisances (parameter set \texttt{2026.1-USO}). Red dashed: the
single-epoch demodulation of the data with the potentials quoted by each
experiment (the treatment of previous \GSF{} releases). The two
reconstructions differ by about $65$\,MV; the three treatments converge
above $\sim 10$\gv{}. Below $\sim 2$--$3$\gv{} the O$^{*}$ and Fe$^{*}$
splines are not directly constrained by data and the comparison reflects the
spline edge.}
\label{fig:phisource}
\end{figure}

The Ghelfi--Maurin--Derome series is adopted as the default potential
because the data mildly prefer it over the Usoskin~2017
reconstruction~\cite{Usoskin:2017cli}, which runs about $65$\,MV lower
over our windows; the two fits are shipped as the default parameter set
\texttt{2026.1} and the alternative \texttt{2026.1-USO}.
Figure~\ref{fig:phisource} compares the fitted interstellar spectra under
the two reconstructions and under the single-epoch demodulation with the
potentials quoted alongside the data, the treatment of previous \GSF{}
releases. The reconstruction choice acts as a coherent systematic on the
low-rigidity anchor: the interstellar spectrum fitted under the Usoskin
potential lies $10$--$14$\% lower below $2$\gv{}, yet re-modulating each
spectrum with its own potential returns fluxes at Earth that agree to
better than $0.5$\%. The choice affects only the interstellar spectrum
below $\sim 10$\gv{}, which is therefore conditioned on the adopted
modulation model and the force-field approximation, and is
self-consistent under them.

\subsection{Uncertainty propagation}

The covariance $C$ of the fitted amplitudes propagates to any derived
quantity $g$ through its Jacobian, $\Sigma_g = (\partial g/\partial a)\, C\,
(\partial g/\partial a)^{T}$, evaluated analytically for fluxes and moments
or by finite differences otherwise. The deviation of an external flux
$J'$ from the \GSF{} is expressed as
\begin{equation}
n_\sigma^2 = \sum_{i,j}\,(J'_i-J_i)\,\bigl(\Sigma^{-1}\bigr)_{ij}\,(J'_j-J_j),
\label{eq:smnsigma}
\end{equation}
where $i$ and $j$ run over the energies $E_i$ of a grid on which the two fluxes
are compared, $J_i$ is the \GSF{} flux at $E_i$, and
$\Sigma_{ij}=\mathrm{Cov}(J_i,J_j)$ is the \GSF{} flux covariance on that same
grid, obtained from the Jacobian rule above with $g$ the vector of fluxes at
those energies. Since $\Sigma$ inherits the rank of $C$, the grid is taken
coarse enough for it to be invertible. Using the full matrix rather than its
diagonal is what makes $n_\sigma$ account for the correlations across energy. Because statistical and
systematic uncertainties are combined in $C$, such $n_\sigma$ values order
deviations consistently but do not carry strict frequentist coverage.

\section{Detailed data treatment}
\label{sm:datatable}

Table~\ref{tab:treatment} summarizes the per-experiment treatment behind
Table~I of the main text: the observation windows used for the solar
modulation of direct data, the hadronic interaction models underlying
air-shower results, and the covariance treatment.

\begin{table*}
\caption{Detailed treatment of the data sets in the fit. Windows apply to
the solar-modulation average of direct data, as published, and are the windows
actually used in the modulation average; the hadronic-model column lists the
interpretation underlying air-shower results as used here. The
$\delta\phi$ column gives the fitted solar-modulation nuisance shift of
each window (Sec.~\ref{sm:solarmod}) with its posterior uncertainty and,
in parentheses, the prior width; the common mode shared by all modulated
data sets fits to $\delta\phi_0=-5\pm22\,(26)$\,MV.}
\label{tab:treatment}
\begin{ruledtabular}
\begin{tabular}{p{2.3cm}p{3.7cm}p{3.4cm}cp{4.3cm}}
\raggedright Experiment & \raggedright Window / model & \raggedright Error treatment & $\delta\phi$ [MV] (prior) & \raggedright Notes \tabularnewline
\colrule
\raggedright ACE-CRIS & \raggedright 2009/03--2010/01 & \raggedright stat $\oplus$ sys & $-72\pm14\;(30)$ & \raggedright solar-minimum window; sub-leading elements \tabularnewline
\raggedright HEAO-3 & \raggedright 1979/10--1980/06 & \raggedright stat $\oplus$ sys & $+49\pm22\;(30)$ & \raggedright heavy elements P--Ni \tabularnewline
\raggedright PAMELA & \raggedright 2006/07--2008/03 and --2008/12 & \raggedright stat $\oplus$ sys & $-19\pm9\;(21)$ & \raggedright deepest solar minimum \tabularnewline
\raggedright AMS-02 & \raggedright 2011/05--2018/05, --2019/10, --2021/05 per element & \raggedright stat $\oplus$ sys & $+5\pm9\;(10)$ & \raggedright D from dedicated release \tabularnewline
\raggedright CALET & \raggedright 2015/10--2022/04 per element; Cr, Ti to 2023/11 & \raggedright stat $\oplus$ sys & $+1\pm11\;(11)$ &  \tabularnewline
\raggedright DAMPE & \raggedright 2016/01--2025/01 & \raggedright stat $\oplus$ sys & $0\pm10\;(10)$ & \raggedright one joint offset (incl.\ boron) \tabularnewline
\raggedright ISS-CREAM & \raggedright 2017/08--2019/02 & \raggedright stat $\oplus$ sys & $0\pm24\;(24)$ &  \tabularnewline
\raggedright NUCLEON-KLEM & \raggedright 2015/07--2017/06 & \raggedright stat $\oplus$ sys & $0\pm22\;(22)$ &  \tabularnewline
\raggedright HAWC & \raggedright QGSJET-II-04 & \raggedright stat $\oplus$ sys & --- &  \tabularnewline
\raggedright GRAPES-3 & \raggedright QGSJET-II-04 & \raggedright stat $\oplus$ sys & --- & \raggedright no quoted scale systematic; 25\% assigned (fit insensitive: preferred scale $f_e\simeq\gsfEscaleGrapes$ at $\gsfEscaleZGrapes\sigma$) \tabularnewline
\raggedright LHAASO & \raggedright mean of QGSJET-II-04, EPOS-LHC, SIBYLL-2.3d & \raggedright stat $\oplus$ sys $\oplus$ half model spread & --- & \raggedright joint p--He block from light elem. sample \tabularnewline
\raggedright IceCube (coincident surface + in-ice) & \raggedright SIBYLL 2.1 & \raggedright per-bin $4\times4$ fraction correlations & --- & \raggedright published covariance \tabularnewline
\raggedright IceTop (low-$E$) & \raggedright SIBYLL 2.1 & \raggedright stat $\oplus$ sys & --- & \raggedright own energy-scale offset: different trigger, reconstruction, years \tabularnewline
\raggedright Tunka-133 & \raggedright QGSJET-II-04 & \raggedright stat $\oplus$ sys & --- &  \tabularnewline
\raggedright TALE (hybrid) & \raggedright QGSJET-II-04 & \raggedright stat $\oplus$ sys & --- & \raggedright $\mlna$ only \tabularnewline
\raggedright Telescope Array & \raggedright --- & \raggedright stat $\oplus$ sys & --- & \raggedright SD, $E\ge 10^{18.25}$\,eV \tabularnewline
\raggedright Pierre Auger & \raggedright SIBYLL-2.3e and EPOS LHC-R (main text) & \raggedright posterior correlations & --- & \raggedright covariance computed from published trials \tabularnewline
\end{tabular}
\end{ruledtabular}
\end{table*}

\subsection{CALET elemental spectra versus the ensemble}
\label{sm:calet}

\begin{figure*}
  \includegraphics[width=\textwidth]{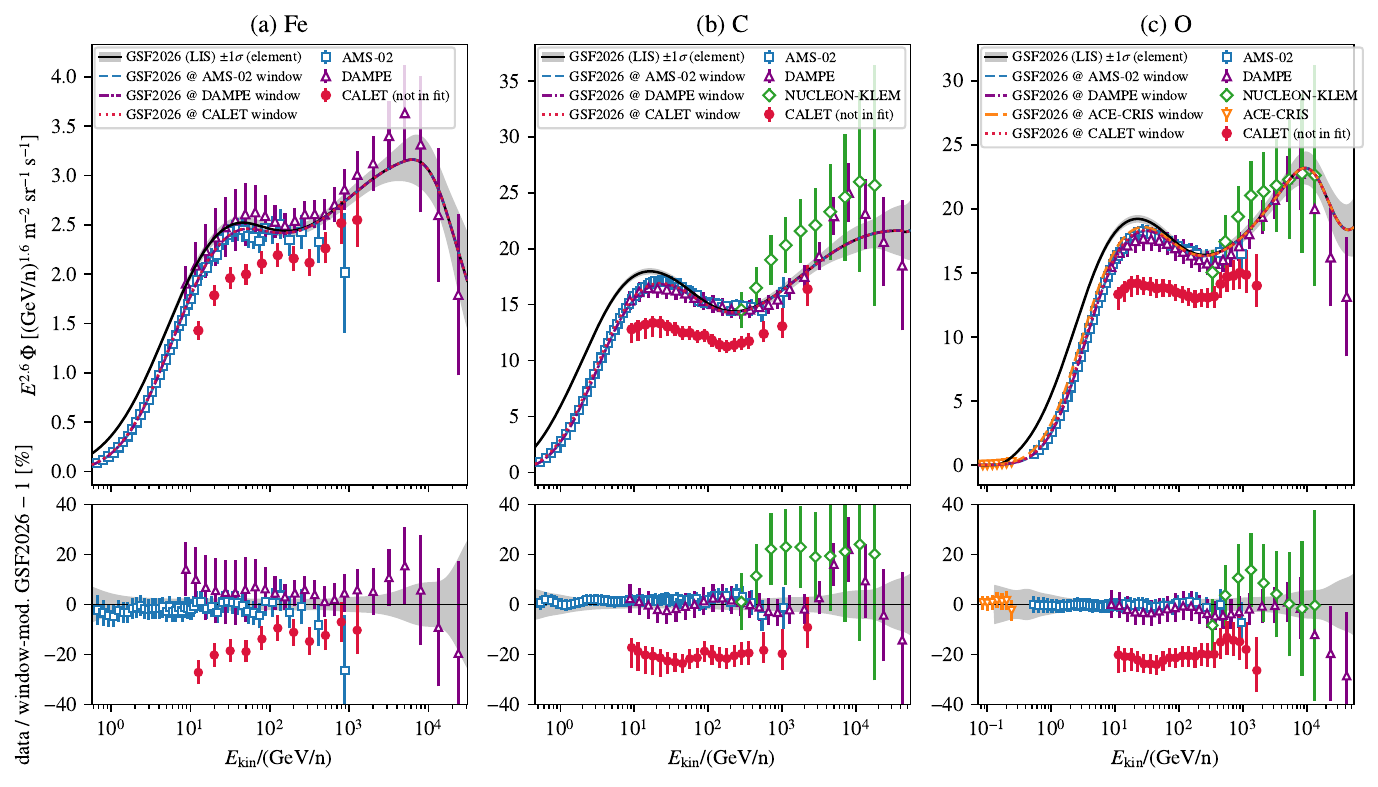}
  \caption{Fitted single-element fluxes of Fe, C, and O ($1\sigma$ bands)
  compared with the direct measurements. Data sets in the fit (open markers)
  are shown on the fitted energy scales; each is compared with the \GSF{}
  model forward-modulated over that experiment's observation window (colored
  curves; the unmodulated LIS in black). The CALET spectra not considered
  in the fit (filled
  red) sit $15$--$20\%$ below the ensemble. Lower panels: relative residuals
  to the window-modulated model. The ACE-CRIS oxygen points at
  $0.09$--$0.25$\gev{}/n (2009 solar-minimum window) follow their modulated
  curve to better than $0.6\sigma$.}
  \label{fig:smcalet}
\end{figure*}

The CALET selection in Table~I of the main text is deliberately asymmetric:
the proton, helium, nickel, titanium, and chromium spectra enter the fit,
while the carbon, oxygen, and iron spectra do not.
Figure~\ref{fig:smcalet} documents the basis for this choice: for each of
Fe, C, and O it compares the fitted single-element flux and its $1\sigma$
band with all direct measurements, each drawn against the model
forward-modulated over that experiment's observation window. The data sets
in the fit are mutually consistent, with no point pulling beyond
$2\sigma$; for iron, where AMS-02 and DAMPE disagree in normalization, the
fit settles between them and the de-weighting widens the band accordingly.
The CALET spectra of the same elements, by contrast, sit $15$--$20\%$
\emph{below} this ensemble with no significant shape difference (mean
pulls $-2.8$ to $-4.5\sigma$). A discrepancy of this size cannot be absorbed
by the CALET energy-scale offset (2\% quoted) and would be inherited by
the covering band of every affected element; the C, O, and Fe spectra are
therefore not considered. The tension is documented by the CALET collaboration
itself for iron against AMS-02~\cite{CALET:2025dgy} and is under
investigation.

The offset is not a single global CALET normalization: the CALET proton
and helium spectra agree with the spectrometer anchor within the fitted
scale offset, and the titanium and chromium spectra agree with the
HEAO-3-anchored sub-iron elements. The published Ti/Fe and Cr/Fe
\emph{ratios} of Ref.~\cite{CALET:2025dgy} are consistent with HEAO3-C2,
since ratios cancel a common normalization; the absolute-flux comparison
shown here is what discriminates.

\subsection{Low-energy band coverage}
\label{sm:lowe}

\begin{figure}
  \includegraphics[width=\columnwidth]{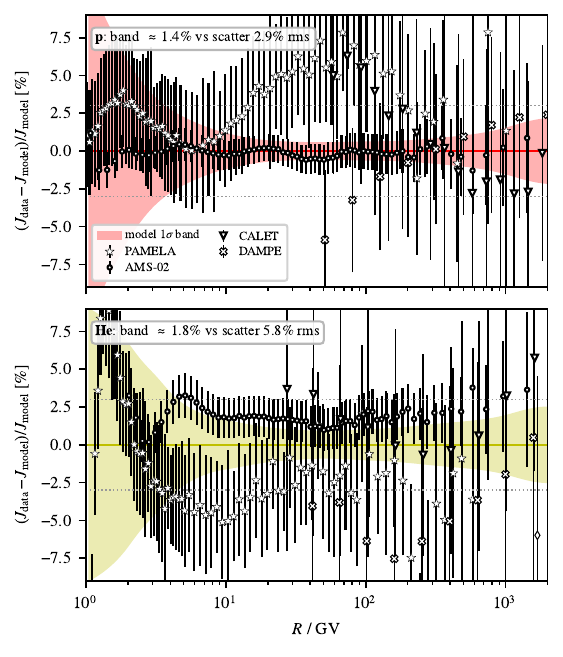}
  \caption{Low-energy proton and helium measurements, shown as measured at
  Earth against the \GSF{} model forward-modulated over each experiment's own
  observation window with that experiment's fitted modulation potential. The
  shaded band is the model's local-interstellar $1\sigma$ band at the plotted
  rigidity. Above $\sim 3$\gv{}, where the band is set by the AMS-02
  statistics, it is narrower than the inter-experiment scatter, which sets the
  effective low-energy uncertainty; below a few \gv{} the band is instead
  dominated by the solar-modulation nuisance (Sec.~\ref{sm:solarmod}).}
  \label{fig:lowe}
\end{figure}

At the lowest rigidities the \GSF{} band has two contributions: a
statistical component of order 1\%, set by the AMS-02 precision, and the
solar-modulation nuisance (Sec.~\ref{sm:solarmod}), which dominates below
a few \gv{} and brings the total to about 7\% for protons and helium at
$1.5$\gv{}, converging to the statistical floor above $\sim 20$\gv{}.
This is comparable to the scatter between the experiments themselves:
measured against the model forward-modulated over each experiment's own
observation window, the proton and helium
measurements scatter with an rms of $3$--$6$\%, dominated by normalization
differences between the calorimetric instruments and AMS-02
(Fig.~\ref{fig:lowe}). The band therefore covers the inter-experiment
spread where the modulation nuisance dominates; the residual normalization
scatter enters the fit through the per-bin de-weighting described in the
main text.

\section{Reduced representation of the covariance}
\label{sm:reduced}

The released model contains the full covariance $C$ of the fitted spline
amplitudes: 85 nonzero leading amplitudes (of 107, the remaining 22 being
zero-padded coefficients held at the NNLS boundary), and, because the published
set is the covering mixture, entries for the sub-leading amplitudes as well.
The latter are populated by the rank-one between-model term of Eq.~(8) of
the main text, which is nonzero wherever the two hadronic interpretations
differ; in a single-interpretation fit only the leading amplitudes carry
covariance entries. All
bands in this work are propagated directly from $C$. The
15 fitted energy-scale offsets are absent because the published amplitude
block is already marginalized over them; their contribution to the flux
uncertainty is therefore contained in $C$.

\begin{figure}
\includegraphics[width=\columnwidth]{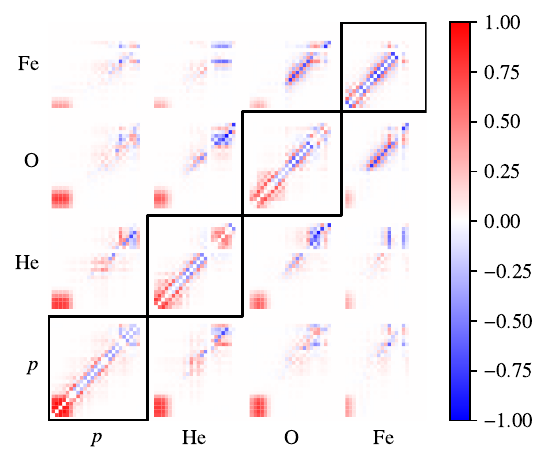}
\caption{Correlation matrix of the fitted spline amplitudes, grouped into
H$^{*}$, He$^{*}$, O$^{*}$, and Fe$^{*}$ blocks.}
\label{fig:cov}
\end{figure}

For use as nuisance parameters in downstream fits, such as
atmospheric-lepton calculations or detector analyses, and for Monte Carlo
propagation, the released model also contains a compact representation of
this uncertainty. It serves the same role as the principal-component
parametrization of the \GSF{} uncertainty employed in
daemonflux~\cite{Yanez:2023lsy}; here each parameter is the relative
deviation of the flux at one fixed energy, rather than an eigenmode
amplitude, and is therefore directly interpretable. The representation
deforms the total proton and neutron
fluxes of Sec.~VI of the main text relative to the central model,
\begin{equation}
J_s(E;\boldsymbol{\theta})=J_{s,\mathrm{central}}(E)
\Bigl[1+\sum_{k=1}^{12}H_k(E)\,\theta_{s,k}\Bigr],
\quad s\in\{\mathrm{p},\mathrm{n}\},
\label{eq:smreduced}
\end{equation}
where the $H_k$ are cardinal interpolation functions in $\log E$ on twelve
pivot energies $E_k$: local cubic splines with $H_k(E_j)=\delta_{jk}$ that
sum to one, each confined to the two intervals around its pivot, with
the edge deformation held constant outside the pivot range. Each component
$\theta_{s,k}$ is therefore the relative deviation of the flux of species
$s$ at the energy $E_k$, and $\boldsymbol{\theta}=0$ returns the central
model. The covariance of the 24 components is evaluated exactly from the
full amplitude covariance,
\begin{equation}
C_\theta = J_{\rm rel}\,C\,J_{\rm rel}^{T},
\label{eq:smpivotcov}
\end{equation}
with $J_{\rm rel}$ the Jacobian of the relative fluxes with respect to the
spline amplitudes, evaluated at the pivot energies. At the pivots the
representation is exact: the variances and all proton--neutron and
pivot-to-pivot correlations equal those of the full model. A downstream fit varies one
component at a time to build the Jacobian of its observable and constrains
the components with the Gaussian penalty
$\chi^2_{\rm prior}=\boldsymbol{\theta}^{T}C_\theta^{-1}\boldsymbol{\theta}$.
Correlated samples $\boldsymbol{\theta}\sim\mathcal{N}(0,C_\theta)$ can
also be drawn; each draw is a smooth deformation of the proton and neutron
fluxes within the model uncertainty.

\begin{figure}
\includegraphics[width=\columnwidth]{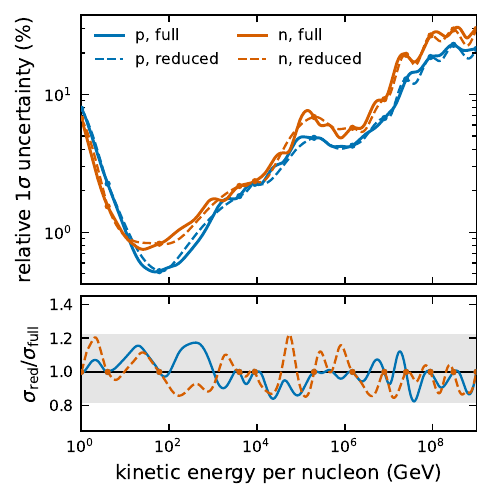}
\caption{Pivot components versus the full covariance. Top: relative
$1\sigma$ uncertainty of the total proton and neutron fluxes from the full
amplitude covariance (solid) and from the 24 pivot components (dashed);
dots mark the pivot energies. Bottom: ratio of the two; within the shaded
band the reduced uncertainty stays inside the worst-case coverage factor of
1.22, and it is exact at the pivots.}
\label{fig:reduced}
\end{figure}

The pivot grid is part of the model definition and is distributed with each
parameter set. The twelve energies minimize the worst-case mismatch between
the reduced and the exact standard deviation,
$\max|\ln(\sigma_{\rm red}/\sigma_{\rm full})|$ over both species and a
dense energy grid, and are rounded; for the 2026 set they are 1, 4, and
60\gev{}, 4, 9, and 200~TeV, 1.5, 8, 25, 90, and 300~PeV, and 1~EeV per
nucleon. Figure~\ref{fig:reduced} compares the reduced and the full
uncertainty: the two agree exactly at the pivots and remain within the
worst-case factor of 1.22 of each other everywhere between them, with no
systematic bias in either direction. Since the grid is optimized for each
parameter set separately, components of different sets refer to different
pivot energies. The grid can also be supplied by the user, for instance to
concentrate the pivots in the energy range of a downstream application; the
distributed grids serve as a starting point.

The pivot components describe the total proton and neutron fluxes on
$1$--$10^{9}$\gev{} per nucleon, which covers atmospheric-lepton
applications, and are the recommended representation for Jacobian-based
propagation of these nucleon fluxes.

\begin{figure}
\includegraphics[width=\columnwidth]{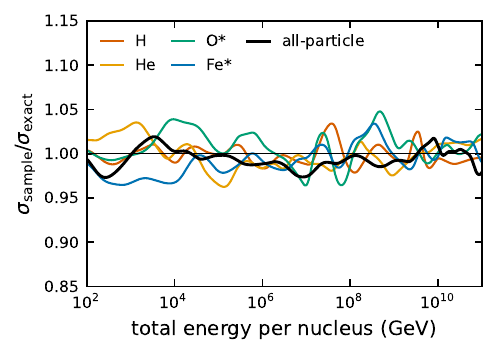}
\caption{Sampling check for pseudo-experiments drawn from the amplitude
covariance: empirical standard deviation of 2000 draws divided by the exact
Jacobian-propagated uncertainty, for the four mass-group fluxes and the
all-particle flux, in total energy per nucleus. The residual scatter is the
statistical fluctuation of the ensemble.}
\label{fig:nativesample}
\end{figure}

When the fluxes of individual nuclei are needed, or when their correlations
with the all-particle flux matter, we recommend pseudo-experiments drawn
from the amplitude covariance instead. The flux is linear in the spline
amplitudes, so a draw from $C$ evaluated with the model reproduces the
covariance at every energy and retains the cross-group correlations
(Fig.~\ref{fig:nativesample}); the
group fluxes of each draw sum to its all-particle flux. The released
package provides such draws through a single \texttt{sample} call on every
model class. As in any Gaussian linear model, sampled fluxes can fluctuate
below zero; this is confined to individual group fluxes in the data-free
tails, where relative uncertainties reach unity.

\section{Fit diagnostics}
\label{sm:diagnostics}

\begin{figure}
  \includegraphics[width=\columnwidth]{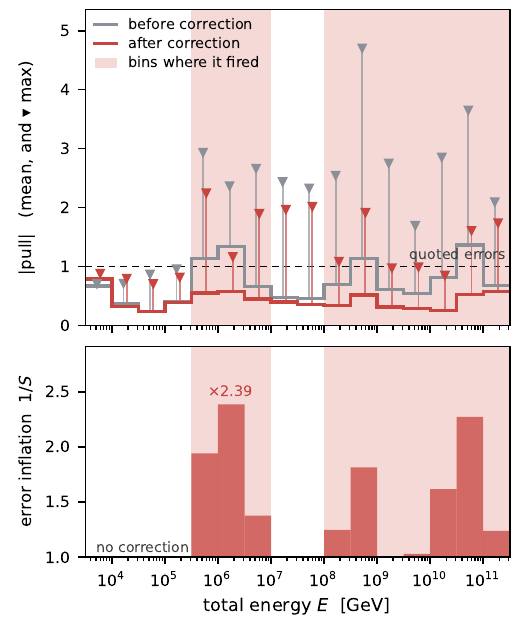}
  \caption{The two-pass weight correction. The passes act on disjoint data and
  are binned in different variables, mapped here onto a common kinetic energy:
  pass~1 treats the flux points of each leading element separately (rigidity
  bins, converted per element), pass~2 treats the all-particle spectra, the mass
  fractions and $\mlna$ together (total-energy bins). Upper, for pass~2: the
  mean $|$pull$|$ per bin (steps) and the largest pull in that bin
  ($\blacktriangledown$), before (gray) and after (red); the dashed line is what
  the quoted errors would give, and shading marks the bins where the correction
  triggered. Lower: the applied error inflation $1/S$, with
  $S=1/\sqrt{\chi^2_\nu}$ of the bin. Pass~2 triggers a correction in nine bins,
  where the quoted uncertainties do not cover the inter-experiment spread: the
  knee, the IceCube/IceTop--Tunka overlap and the Auger-versus-Telescope-Array
  bins. There it lowers the mean pull from $0.91$ to $0.45$ and the largest from
  $4.6$ to $2.2$. Pass~1 triggers in fourteen bins, most strongly for helium
  below $4$\gev{} ($\chi^2_\nu=10.4$, inflation $3.2$), where AMS-02 and PAMELA
  disagree.}
  \label{fig:coveringcomp}
\end{figure}

\begin{figure*}
  \includegraphics[width=\textwidth]{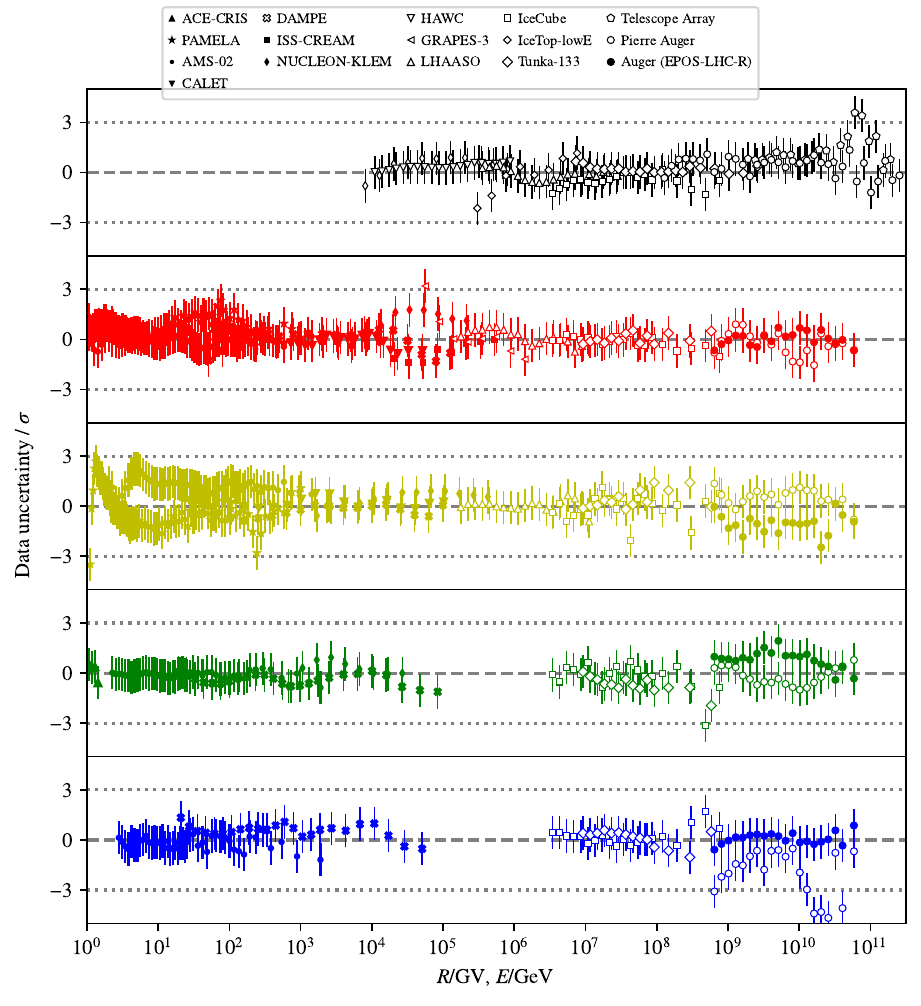}
  \caption{Pulls (data $-$ model)/$\sigma$ of all fitted data sets on the
  common energy scale, using the same reference as the likelihood:
  solar-modulated measurements are pulled against the forward
  window-average-modulated model (Sec.~\ref{sm:likelihood}), all others
  against the interstellar flux. The Auger fluorescence fractions are shown
  under both hadronic-interaction models: open circles denote the fitted
  SIBYLL-2.3e interpretation, filled circles the EPOS-LHC-R overlay (not
  fitted), both pulled against the same model.}
  \label{fig:pulls}
\end{figure*}

\begin{figure*}
  \includegraphics[width=\textwidth]{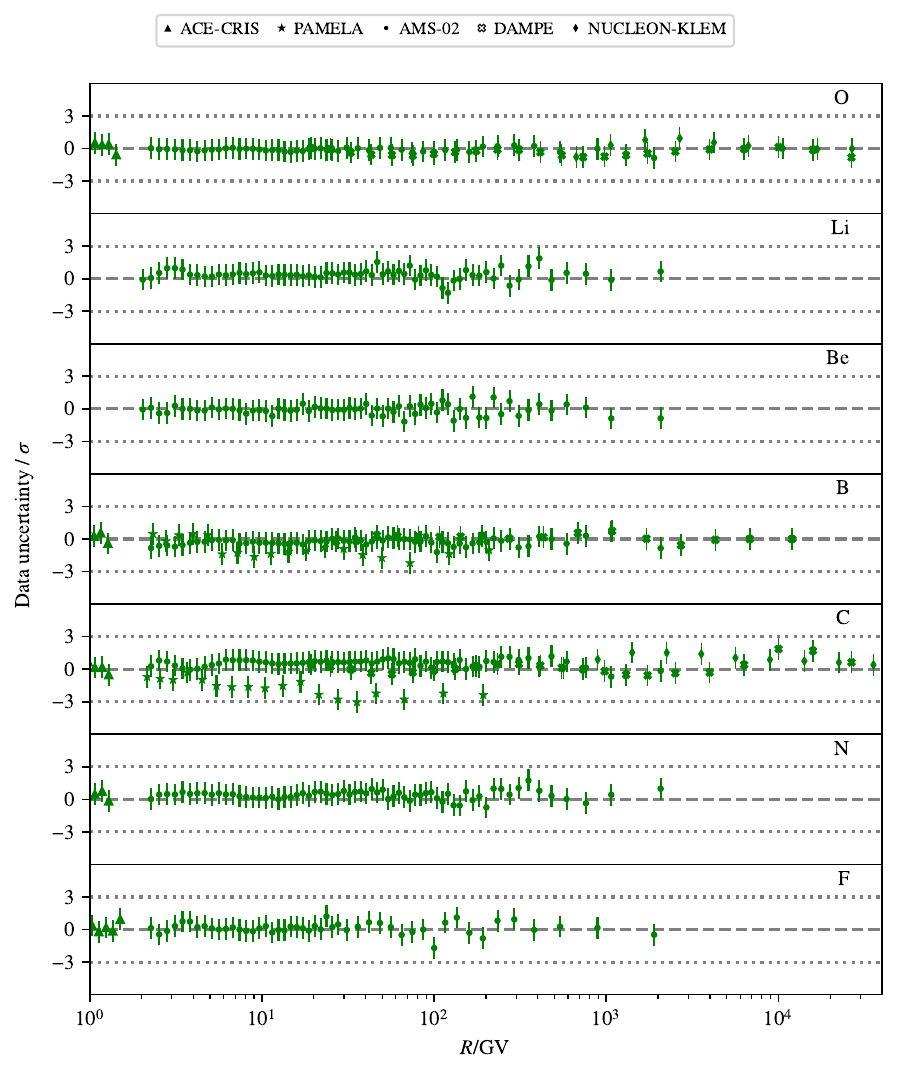}
  \caption{Pulls of the oxygen-group observables.}
  \label{fig:pullsO}
\end{figure*}

\begin{figure*}
\includegraphics[height=0.92\textheight]{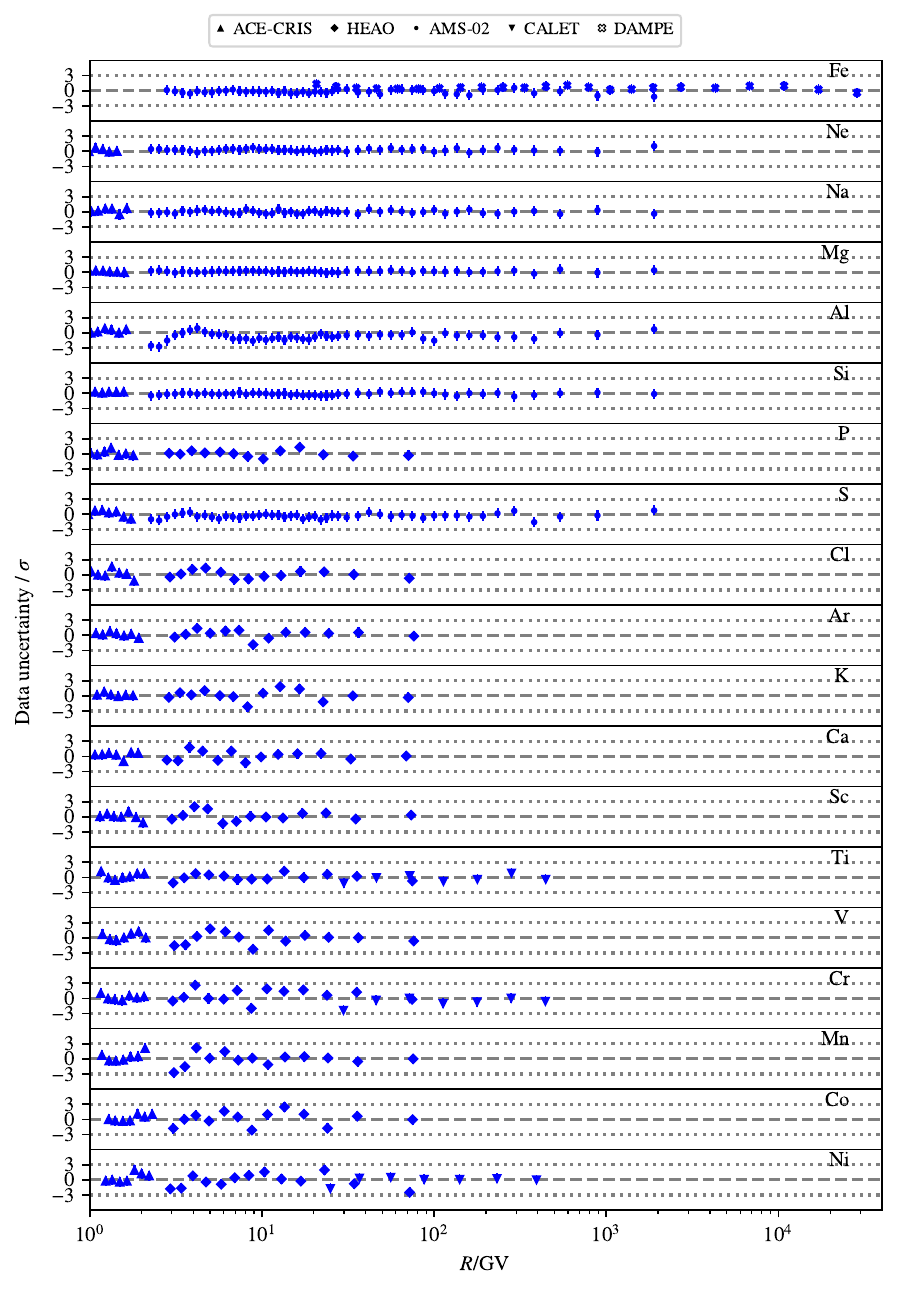}
\caption{Pulls of the iron-group observables.}
\label{fig:pullsFe}
\end{figure*}
\begin{figure*}
\includegraphics[width=\textwidth]{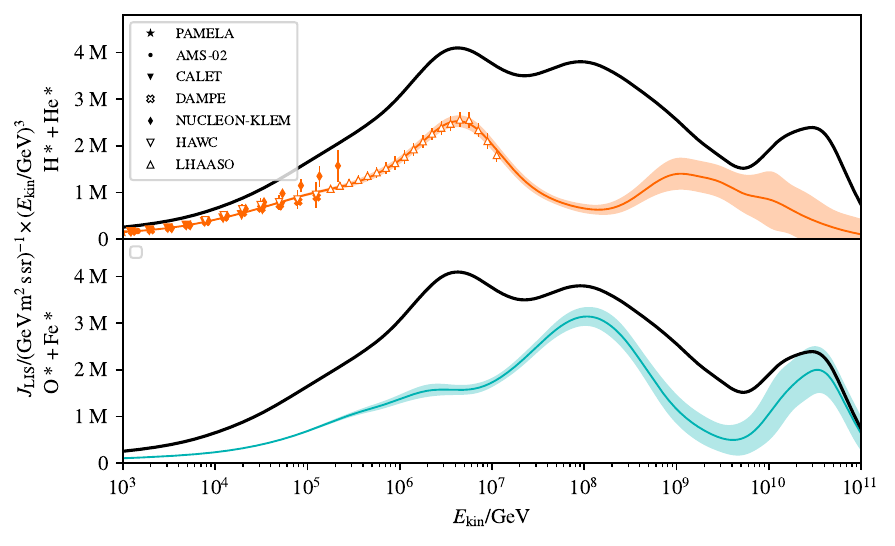}
\caption{Light (H$^{*}$+He$^{*}$) and heavy (O$^{*}$+Fe$^{*}$) fluxes for experiments
publishing two-group splits, compared with the \GSF{}.}
\label{fig:2frac}
\end{figure*}

Figure~\ref{fig:pulls} shows the pulls of all fitted data sets against the
model on the common energy scale; the group-resolved panels for the oxygen
and iron groups are shown in Figs.~\ref{fig:pullsO} and \ref{fig:pullsFe}.
The pull distributions are consistent with the reduced $\chi^2$ of the fit
and show no systematic trends with energy beyond the tensions discussed in
the main text: the mean pull of the direct data stays within $+0.1$ to
$+0.3\sigma$ in every rigidity decade. The only points beyond $5\sigma$ are
the two lowest PAMELA helium points at $1.0$--$1.1$\gv{}, where that spectrum
turns over by tens of percent within a single decade of rigidity. The feature
is far too steep to be a modulation-potential effect and is not reproduced by
the AMS-02 helium data over the same interval, so it is treated as a PAMELA
systematic rather than a deficiency of the fit.
Figure~\ref{fig:2frac} shows the light/heavy decomposition used by
experiments that publish two-group splits.

Figure~\ref{fig:coveringcomp} shows the two-pass weight correction (main
text, Sec.~II.D) at work. The second pass triggers a correction only in the
total-energy bins where the ensemble genuinely disagrees: the knee, the
IceCube/IceTop--Tunka overlap, and the Auger-versus-Telescope-Array bins, with
reduced $\chi^2$ up to about $5$. There it inflates the quoted
uncertainties by $S=1/\sqrt{\chi^2_\nu}\simeq0.5$--$0.8$, bringing the
maximum per-bin pull down from $2.5$--$4.7$ to below $2.2$ while leaving
the consistent bins untouched. The re-weighting also shifts the central
all-particle flux by a few percent at $10^{6}$--$10^{9}$\gev{} through a
smooth re-equilibration of the energy-scale offsets; the single-pass
central value lies within the widened band. The single-pass fit remains
available in the fitter and is reproduced bit-identically by its closure
test.

\clearpage

\end{document}